\documentclass[journal,onecolumn,12pt,draftclsnofoot]{IEEEtran}
\usepackage{graphicx,psfrag,epsfig,epsf,latexsym,hhline,amsmath,amssymb,multirow}

\usepackage[usenames,dvipsnames]{pstricks}
\usepackage{pst-plot}
\usepackage{caption}
\usepackage{xcolor}
\usepackage{subcaption}
\usepackage{enumitem} 
\usepackage{amsmath} 
\usepackage{color}
\usepackage{hyperref}
\usepackage{bbm}
\usepackage{graphicx}
\usepackage{amsthm}
\usepackage{algorithm}
\usepackage{algpseudocode}
\usepackage[noadjust]{cite}
\usepackage{blindtext}
\usepackage{etoolbox,tcolorbox}
\usepackage{stfloats}
\usepackage{tikz}
\usetikzlibrary{chains,shapes.multipart}
\usetikzlibrary{shapes,calc,fit}
\usetikzlibrary{automata,positioning}

\newcounter{cntr}
\tikzset{
    queuei/.pic={
  \stepcounter{cntr}
        \node[outer sep=0pt,draw,rectangle split,rectangle split horizontal,minimum height=10pt,rectangle split parts=3] (queue-\thecntr) [pic actions] {};
        \draw
          (queue-\thecntr.north west) -- ++(-10pt,0)
          (queue-\thecntr.south west) -- ++(-10pt,0);
    },
}

\input{Jerry.def}

\begin{document}
\title{Scheduling for Periodic Multi-Source Systems with Peak-Age Violation Guarantees}
\author{Kuan-Yu Lin$^\dag$, Yu-Chih Huang$^\dag$, and Yu-Pin Hsu$^*$\\
\small $^\dag$Institute of Communications Engineering, National Yang Ming Chiao Tung University, Hsinchu, Taiwan\\
$^*$Department of Communication Engineering, National Taipei University, New Taipei City, Taiwan\\
E-mail: \{casperlin.ee09@nycu.edu.tw, jerryhuang@nycu.edu.tw, yupinhsu@mail.ntpu.edu.tw\} }

\maketitle

\begin{abstract}
Age of information (AoI) is an effective performance metric measuring the freshness of information and is particularly suitable for applications involving status update. In this paper, using the age violation probability as the metric, scheduling for heterogeneous multi-source systems is studied. Two queueing disciplines, namely the infinite packet queueing discipline and the single packet queueing discipline, are considered for scheduling packets within each source. A generalized round-robin (GRR) scheduling policy is then proposed to schedule the sources. 
Bounds on the exponential decay rate of the age violation probability for the proposed GRR scheduling policy under each queueing discipline are rigorously analyzed.
Simulation results are provided, which show that the proposed GRR scheduling policy can efficiently serve many sources with heterogeneous arrivals and that our bounds can capture the true decay rate quite accurately. When specialized to the homogeneous source setting, the analysis concretizes the common belief that the single packet queueing discipline has a better AoI performance than the infinite packet queueing discipline. Moreover, simulations on this special case reveals that under the proposed scheduling policy, the two disciplines would have similar asymptotic performance when the inter-arrival time is much larger than the total transmission time.

\end{abstract}




\section{Introduction}\label{sec:intro}
The success of many new applications in the Internet of Things (IoT) hinges greatly on the timeliness of information. An outstanding example is the industry 4.0 where stringent constraints on latency are usually required in order to maintain the sophisticated collaboration among distributed devices and factories. To address timeliness of information, the age of information (AoI) was introduced in \cite{Kaul12_age_provide} as a new performance metric to measure the freshness of information. AoI measures the amount of time elapsed since the generation of the latest updated packet until the present, which is fundamentally different from the notion of delay and is arguably a better performance metric than throughput and delay considering timeliness of information \cite{Sun19}. It has also been demonstrated in \cite{Yates21} that optimal design based on delay or throughput criteria does not necessarily minimize the AoI. For example, to increase the throughput, one may want to backlog the packets in the queue, which inevitably updates stale information more often. On the other hand, to decrease the delay, one may reduce the update rate, which in turn generates fresh data less frequently.

Despite many successes in the AoI literature thus far, most of them focused solely on using the average AoI as the metric. This includes an intensive study on analyzing the AoI of a signle source network for a various queueing disciplines such as the last come first serve (LCFS) queue \cite{Kaul12_LCFS}, the G/G/1 queue with the first come first serve (FCFS) and LCFS in \cite{Soysal21}, and M/M/1, M/M/2, and M/M/$\infty$ systems in \cite{Kam16}.  Although minimizing average AoI more or less implies reducing the probability of violating the AoI threshold, it cannot be directly converted to strict performance guarantees such as the age violation probability. Therefore, these results are not directly applicable to applications that have strict age requirements. To cope with this, there were a series of works \cite{Jaya19,Jaya21_multi_hop,Limei21,Inoue19,Seo19, Song21} analyzing the distribution of AoI. In \cite{Jaya19}, the authors provided the age violation distribution of the single source single server system under D/GI/1 and M/GI/1 queuing systems. In \cite{Jaya21_multi_hop}, Champati {\it et al.} provided a upper bound of age violation probability under multi-hop systems. In \cite{Limei21}, the age violation probability for the M/M/1 and M/D/1 one source-server pair systems was analyzed. Since these systems considered by the above works have only one queue, the transmission schedule is fully determined by the underlying queueing discipline and there does not require extra scheduling among sources.\footnote{In this work, we use the term ``queueing discipline" to describe how packets are scheduled within the same source and the term ``scheduling" to specify how packets of different sources are scheduled among the sources.} Inoue {\it et al.} in \cite{Inoue19} derived the stationary distribution of AoI in terms of the stationary distribution of delay and peak AoI for various queueing disciplines. In \cite{Seo19}, Seo and Choi provided the performance guarantee for the single source system with FCFS queueing discipline by analyzing an upper bound of the age violation probability as the threshold goes to infinity. The authors of \cite{Song21} discussed the benefit of retransmission and provided the closed-form expression of age violation probability in single source-destination pair. However, these works focused solely on the single source system and the analysis therein again are not directly applicable to the multi-source scenario.

For multi-source/multi-user networks with homogeneous applications/devices, scheduling among sources has been studied. With average AoI as the metric, scheduling has been investigated in \cite{Hsu20,Hsu18,Kadota18,Maatouk21,Jiang19,Moltafet20,Kumar23}. In \cite{Hsu20}, it was shown that the optimal scheduling algorithm is stationary and deterministic; also the asymptotically optimal scheduling policy for multi-user systems with stochastic arrival was provided. In \cite{Hsu18}, a structural Markov decision process (MDP) scheduling algorithm and an index scheduling algorithm were proposed and fully analyzed. In \cite{Kadota18}, three low-complexity scheduling policies were considered and analyzed, namely the randomized policy, the Max-Weight policy and the Whittle's index policy. In \cite{Maatouk21},  the Whittle's index policy was further proved to minimize the average AoI in the multi-source regime. In \cite{Jiang19}, the round robin (RR) policy was investigated under the queueing discipline that only keeps the latest packet and it was shown that such the simple policy is in fact optimal among all arrival-independent policies. 
In \cite{Moltafet20}, Moltafet {\it et al.} analyzed the average AoI for the multi-soruce network with M/M/1 (respectively M/G/1) system under FCFS queueing discipline. In \cite{Kumar23}, the same setting was considered with exception that the server may break down in order to study the effect of service disruptions.

Another important feature of IoT is that there are typically a number of heterogeneous applications/devices inhabiting the system. For such a scenario, scheduling among sources needs to be carefully designed in order to keep information fresh for each user \cite{Chowdhury18}. In this work, we consider a multi-source system with heterogeneous source types, where packets of different source types arrive periodically with different periods.
This assumption is not merely for easing the analysis, but also to address a practical scenario where resource is pre-allocated in a periodic fashion (see semi-persistent scheduling in NB-IoT systems \cite{Sergey19,Wang17} for example) and applications with higher AoI requirements tend to sample more frequently. On the other hand, the transmission time is stochastic and i.i.d. that can have any distribution as long as the moment generating function is finite. Moreover, two frequently encountered queueing disciplines, namely the infinite packet queueing (IPQ) with FCFS discipline and the single packet queueing (SPQ) discipline, are considered. We propose a novel scheduling policy based on the RR algorithm,\footnote{Due to their simplicity and effectiveness demonstrated in \cite{Jiang19}, RR-type policies are considered solely in this paper.} called the generalized round-robin (GRR) policy. To provide a strong performance guarantee for our proposed GRR policy, its age violation probability that the peak AoI exceeds the prescribed threshold is rigorously analyzed.
Given the existing literature reviewed above, to the best of our knowledge, the present work is the first to provide a closed-form solution/approximation of the distribution of AoI in the multi-source system.\footnote{A similar problem has been studied in \cite{Dogan21}, where a multi-source probabilistically preemptive bufferless M/PH/1/1 queueing system with packet errors is considered. Therein, the distribution of AoI in multi-source system was solved only numerically.}
The main contributions of this paper are summarized as follows:
\begin{enumerate}
    \item For a multi-source system with heterogeneous periodic arrivals, we propose a low-complexity scheduling policy called GRR that generalizes the famous RR policy to accommodate the heterogeneity. To provide performance guarantees, the age violation probability of the proposed policy that the prescribed peak AoI threshold is violated is then analyzed under the two considered queuing disciplines. 

    \item For GRR under the IPQ, an upper bound and a lower bound on the age violation probability for any given threshold are derived. Based on the derived bounds, the asymptotic decay rate when the number of sources tends to infinity is approximated. Simulation results show that our analytical results accurately capture the scaling of the age violation performance. Moreover, an approximation based on our bounds is provided, which indicates that the proposed GRR policy enjoys a desired property that the age violation probability decays faster for sources whose packets arrive more frequently (i.e., have higher AoI requirement).

    \item 
    For GRR under the SPQ, we derive an upper bound on the age violation probability. Although a tight lower bound is lacking, simulation results indicate that the derived upper bound again captures the true scaling quite accurately when the number of sources becomes large. They also indicate that the SPQ can handle some situations where the IPQ would have been overflowed. Moreover, an approximation based on the bound again reveals that the proposed GRR policy has the desired property that the age violation probability decays faster for sources whose packet arrive more frequently. Another implication of our results is that as the total number of sources grows, scaling the arrival period linearly with the number of sources is sufficient to drive the age violation probability vanishing.

    \item The derived bounds are specialized to the homogeneous system, where two conclusions can be drawn: 1) the common belief that the SPQ has a better AoI performance than the IPQ is concretized; and 2) the two queueing disciplines result in the same asymptotic scaling when the inter-arrival time is much larger than the total transmission time.
\end{enumerate}

The rest of the paper is organized as follows. In Section \ref{sec:system model}, we illustrate the network model, the age model, and the problem we study. In Section \ref{sec:FCFS}, we provide the analysis of the age violation probability for the IPQ.
Our analysis for the SPQ is then provided in Section \ref{sec:LCFS}.
In Section \ref{sec:numerical and simulation}, we validate our analysis with simulations and compare the proposed GRR under the two queueing disciplines. Finally, in Section \ref{sec:conclusion}, we conclude the paper.

\subsection{Notation}
Throughout the paper, constants and random variables are written in lowercase and uppercase, respectively, for example, $x$ and $X$. Sets are written in calligraphic letters, for example $\mc{A}$. For a real constant $x$, we use $(x)^+$ to denote $\max\{0,x\}$. For a positive integer $n$, $[n]$ is an abbreviation of the set $\{1, 2, \ldots, n\}$. We use $\mathbbm{1}_{\{\cdot\}}$ to represent the indicator function for the event inside the curly brackets.

\begin{table}[htbp]\caption{Notations}
\begin{center}
\begin{tabular}{| r | p{10cm} |}
\hline
Notation & Description \\
\hline
$n$ & total number of sources \\
$\eta$ & total number of groups \\
$n_g$ & number of sources in group $g$\\
$\alpha_g$ & fraction of sources in group $g$\\
$nb$ & fundamental arrival period\\
$d_g$ & multiple number of frequency of group $g$\\
$\Tilde{d}$ & number of rounds in an iteration\\
$(g,i)$ & source index for source $i$ in group $g$\\
$(g,i,k)$ & packet index for the $k$-th updated packet of source $(g,i)$\\
$U_{g,i}(t)$ & generation time of the latest packet of source $(g,i)$ transmitted to the destination by time $t$\\
$S_{g,i}(k)$ & arrival time of the packet $(g,i,k)$\\
$W_{g,i}(k)$ & waiting time of the packet $(g,i,k)$\\
$D_{g,i}(k)$ & departure time of packet $(g,i,k)$\\
$V_{g,i}(k)$ & transmission time of packet $(g,i,k)$\\
$T_{g,i}(k)$ & total transmission time since transmitting the packet $(g,i,k-1)$ to the beginning of the transmission of packet $(g,i,k)$\\
$\Delta_{g,i}(t)$ & AoI for source $(g,i)$ at time $t$\\
$A_{g,i}(k)$ & peak AoI for packet $(g,i,k)$\\
$\Lambda_v(\theta)$ & log moment generating function of transmission time with parameter $\theta$\\

\hline
\end{tabular}
\end{center}
\label{tab:TableOfNotationForMyResearch}
\end{table}



\section{System Model and Problem Formulation} \label{sec:system model}
In this section, we describe the network model in Section~\ref{subsec:net_model}, followed by the notion of AoI and a description of the problem we attempt to solve in Section~\ref{subsec:AoI}. In Table~\ref{tab:TableOfNotationForMyResearch}, we summarize the notations we use in this section for describling the considered problem.

\subsection{Network model}\label{subsec:net_model}
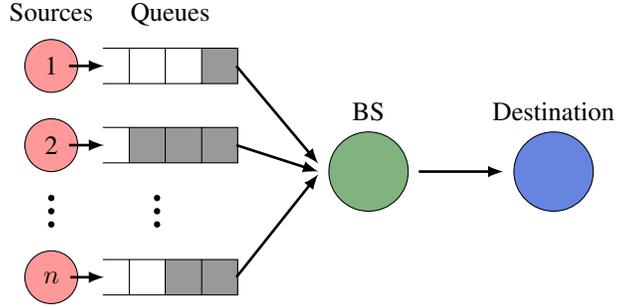
\begin{figure}
\centering
\begin{tikzpicture}[>=latex]
\def\pos{0}
\draw [fill=red!40] (\pos,100pt) circle [radius=10pt] node (s1) {\small 1};
\draw [fill=red!40] (\pos,70pt) circle [radius=10pt] node (s2) {\small 2};
\draw [fill=red!40] (\pos,20pt) circle [radius=10pt] node (sn) {\small $n$};
\foreach \idx in {10, 15, 20}
    \filldraw [black] ([yshift=10+\idx pt]sn) circle (1pt);

\path
(\pos+50pt,100pt) pic[rectangle split part fill={white,white,gray!80}] {queuei=1}
(\pos+50pt,70pt) pic[rectangle split part fill={gray!80,gray!80,gray!80}] {queuei=2}
(\pos+50pt,20pt) pic[rectangle split part fill={white,gray!80,gray!80}] {queuei=n};
\path
(\pos+40pt,100pt) coordinate (Q1)
(\pos+40pt,70pt) coordinate (Q2)
(\pos+40pt,20pt) coordinate (Qn);
\foreach \idx in {10, 15, 20}
    \filldraw [black] ([yshift=10+\idx pt]Qn) circle (1pt);

\draw[line width=1pt, ->] ([xshift=10pt]s1)--([xshift=-20pt]Q1);
\draw[line width=1pt, ->] ([xshift=10pt]s2)--([xshift=-20pt]Q2);
\draw[line width=1pt, ->] ([xshift=10pt]sn)--([xshift=-20pt]Qn);

\draw [fill=black!60!green!50]  (\pos+120pt,60pt) circle [radius=15pt] node{} coordinate (BS);

\draw[line width=1pt, ->] ([xshift=30pt]Q1)--([xshift=-19pt, yshift=3pt]BS);
\draw[line width=1pt, ->] ([xshift=30pt]Q2)--([xshift=-19pt, yshift=0pt]BS);
\draw[line width=1pt, ->] ([xshift=30pt]Qn)--([xshift=-19pt, yshift=-1pt]BS);



\draw [fill=green!20!blue!60] (\pos+190pt,60pt) circle [radius=15pt] node{} coordinate (Dn);

\draw[line width=1pt, ->] ([xshift=19pt]BS)--([xshift=-19]Dn);

\begin{scope}
\node [anchor=south] at ([yshift=5pt]s1.north) {\small Sources};
\node [anchor=south] at ([xshift=5pt, yshift=11pt]Q1.north) {\small Queues};
\node [anchor=south] at ([yshift=15pt]BS.north) {\small BS};
\node [anchor=south] at ([yshift=15pt]Dn.north) {\small Destination};
\end{scope}
\end{tikzpicture}
\caption{An illustration of the network model.}\label{fig:illustrate network}
\end{figure}

We consider an information update system as shown in Fig.~\ref{fig:illustrate network}, where $n$ sources update their respective status of information (in terms of packets) to a destination through a base station (BS).\footnote{The model is equivalent to having $n$ destinations where each source is requested by a destination.}
These $n$ sources are categorized into $\eta$ groups of heterogeneous sources, where group 1 contains the most frequently sampled sources, group 2 contains the second most frequently sampled sources, and so on. We use the pair $(g,i)$ to denote the source $i$ of group $g$. Also, we use $\alpha_g$ and $n_g=\alpha_g n$ for $g\in[\eta]$ to represent the fraction and the number of sources in group $g$, respectively, with the convention $\alpha_0=0$. Periodic sampling is considered, where the sources in the same group have the same arrival period.
Precisely, let $d_g\in\mbb{N}$ with $1=d_1<d_2<\ldots<d_{\eta}$. For $b\in\mbb{R}$ a constant, the arrival period of a source in group $g$ is $d_g nb$, scaled linearly with $n$.\footnote{This is in fact a typical way of analyzing the asymptotic performance in the large-source regime \cite{Srikant14}.} That is, sources with the same sample period arrive in bulk in a deterministic and periodic fashion. We denote by $S_{g,i}(k)$ the arrival time of the $k$-th updated packet of source $(g,i)$.

The BS maintains a queue for each source, where (some of the) unserved packets of source $i\in[n]$ are stored in queue $i$. In this paper, two types of queueing disciplines are considered, namely the IPQ and the SPQ. For the IPQ, the queue size is infinity and all arriving packets are stored and served with FCFS. For the SPQ, to prevent BS from sending stale information, the latest packet is queued and others are preempted.

For each source, the transmission time is random and we use $V_{g,i}(k)$ to denote the transmission time of packet $(g,i,k)$, the $k$-th updated packet of source $(g,i)$. $V_{g,i}(k)$ is assumed to be independent and identically distributed (i.i.d) that has a finite log-moment generating function  $\Lambda_v(\theta) = \log \mathbb{E} \left[e^{\theta V_{g,i}(k)}\right]$. We stress that $V_{g,i}(k)$ can be discrete or continuous random variable; thereby, our model and analysis apply to both the discrete-time and continuous-time systems.

\begin{remark}\label{rmk:d1}
    We can safely assume that $d_1 = 1$. If otherwise, we can always add a virtual group with $d_1=1$ and $V_{1,i}(k)=0$ $\forall i, k$. This will not alter the derivation and analysis in the sequel.
\end{remark}



\subsection{Age of information and problem formulation}\label{subsec:AoI}
\begin{figure}
	\center{\includegraphics[width=2.8in]{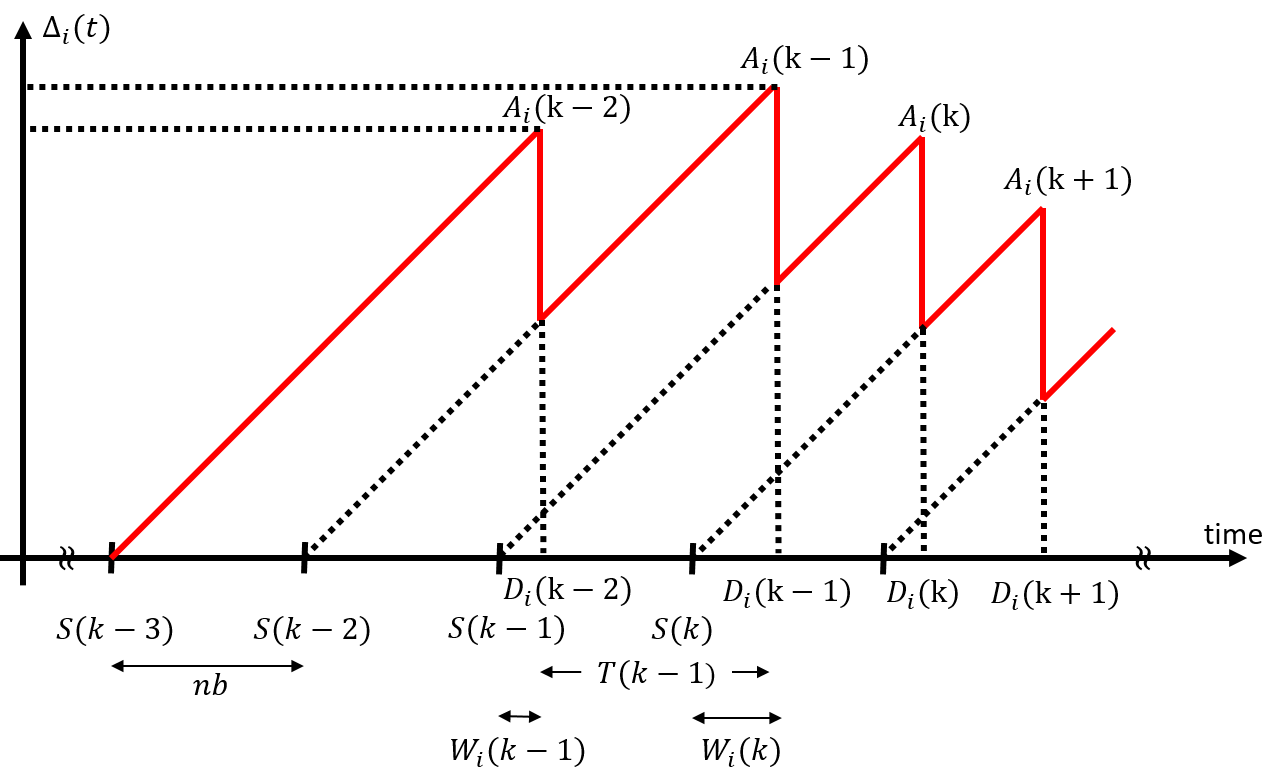}}
   	\caption{ Age evolution in a multi-source system.}
   	\label{fig:age evolution}
\end{figure}

We now describe the definition of AoI. Under scheduling policy $\pi$, let $D^\pi_{g,i}(k)$ be the departure time of the packet $(g,i,k)$ and let $U^\pi_{g,i}(t)$ be the generation time of the latest packet transmitted to the destination by time $t$. (i.e. $U^\pi_{g,i}(t) =  S_{g,i}(k)$ where $k = \arg\;\max_k D^\pi_{g,i}(k)\leq t$) Then, AoI for source $(g,i)$ at time $t$ is defined by $\Delta^\pi_{g,i}(t) = t-U^\pi_{g,i}(t)$ and is depicted in Fig.~\ref{fig:age evolution}. 
The peak age of the packet $(g,i,k)$ is defined as
\begin{equation}\label{eqn:A(k)}
A^\pi_{g,i}(k) = D^\pi_{g,i}(k)-S_{g,i}(k-1),
\end{equation}
which captures the time from the generation of the $(k-1)$-th updated packet of the source $(g,i)$ to the successful departure of the $k$-th updated packet from the same source.

Note that we have used the superscript $\pi$ to emphasize that those random variables depend on the underlying scheduling policy. In this paper, since we focus solely on the generalized round-robin scheduling policy introduced in Section~\ref{subsec:GRR}, we drop the superscript $\pi$ from this point onward. Unlike most of the work in the AoI literature analyzing the long-term average of the peak AoI, to provide performance guarantee and to have a better understanding of the AoI distribution, we aim to analyze the age violation probability defined as follows.

\begin{define}[Age violation probability]
    We define the age violation probability for the packet $(g,i,k)$ as the probability that the packet's peak AoI exceeds a given threshold $nx$. It is expressed as
    $    Pr(A_{g,i}(k) \geq nx).$
\end{define}
Having defined the age violation probability, we then define the asymptotic decay rate, which provides a guarantee on the decay rate of the age violation probability in a large-user system.
\begin{define}[Asymptotic decay rate]
The asymptotic decay rate under the threshold $nx$ is the rate at which the peak age decays as the number of sources $n$ grows to infinity. That is,
\begin{equation}\label{eqn:age requirement}
    \lim_{n \to \infty} \frac{1}{n} \log Pr(A_{g,i}(k) \geq nx).
\end{equation}
\end{define}
In this work, we also aim to analyze the asymptotic age violation probability characterization for our proposed scheduling policy when the number $n$ of sources is large.

\section{GRR under the IPQ}\label{sec:FCFS}

In this section, we focus on the IPQ. We first propose the GRR scheduling algorithm in Section \ref{subsec:GRR}. Next, we analyze the age evolution under GRR together with some insights and explanations in Section \ref{subsec:age analysis FCFS}. Based on the age evolution, we then derive an upper bound and a lower bound on the age violation probability, which allow us to approximate the asymptotic decay rate for the large-system scenario in Section \ref{subsec:asymptotic analysis FCFS}. To gain intuition, we study asymptotic decay rate with a specific distribution for the transmission time in Section \ref{subsec:aprrox_FCFS}.
Moreover,  we specialize our results to the homogeneous case where there is only one single group in \ref{subsec:RR_FCFS}
\subsection{Proposed GRR under the IPQ}\label{subsec:GRR}
Before introducing the proposed GRR scheduling policy, we first present the RR scheduling policy as follows and then define the proposed GRR policy.
\begin{define}[Round-robin policy]\label{def:RR}
Let us define the idle time of source $(g,i)$ at the time $t$ be the time duration since the last time this source was updated to $t$. Then, at the time $t'$ that the previous packet was successfully delivered, the RR policy schedules the source that has the maximum idle time at $t'$.
If there is no packet in the queue of this source, the BS waits until the next arrival of that source.
\end{define}

\begin{define}[Generalized round robin]\label{def:GRR}
    Let $\tilde{d}$ be the least common multiple of $d_1$, $d_2$, ..., $d_{\eta}$. GRR operates iteratively with $\tilde{d}$ rounds in an iteration. In round $r$ such that $r \mmod d_g = 0$, the sources in group $g$ are served according to the RR policy defined in Definition~\ref{def:RR}.
\end{define}
An example is provided in the following.
\begin{example}\label{exp:three_group_system}
Consider $\eta=3$ and $d_1=1,\ d_2=2,\ d_3=3$. There are $\tilde{d}=6$ rounds in an iteration. GRR schedules as follows:
\begin{center}
    \begin{tabular}{c c c c c c}
        Round 0: & $\mc{G}_1$ & $\to$ & $\mc{G}_2$ & $\to$ & $\mc{G}_3$ \\
        Round 1: & $\mc{G}_1$ \\
        Round 2: &$\mc{G}_1$ & $\to$ & $\mc{G}_2$ \\
        Round 3: & $\mc{G}_1$ & $\to$ & $\mc{G}_3$ \\
        Round 4: & $\mc{G}_1$ & $\to$ & $\mc{G}_2$ \\
        Round 5: & $\mc{G}_1$\\
        \hline
        Round 6: & $\mc{G}_1$ & $\to$ & $\mc{G}_2$ & $\to$ & $\mc{G}_3$ \\
        Round 7: & $\mc{G}_1$ \\
        $\vdots$ & $\hspace{25pt}\vdots$
    \end{tabular}
\end{center}
\end{example}

\begin{remark}
For a homogeneous system where there is only one group, the proposed GRR policy reduces to the RR policy, justifying the name. Also, the proposed GRR policy inherits the extreme low-complexity from the RR policy.
\end{remark}

\subsection{Age analysis under the IPQ}\label{subsec:age analysis FCFS}
To analyze the age evolution under the proposed GRR policy, we denote by $T_{g,i}(k-1)$ the total transmission time since transmitting the packet $(g,i,k-1)$ to the beginning of the transmission of packet $(g,i,k)$. $T_{g,i}(k-1)$ includes the transmission time of the $(k-1)$-th update packet of source $i$ but does not include that of the $k$-th updated packet of this source. And it can be expressed as follows:
\begin{equation}\label{eqn:T_g(k)}
    T_{g,i}(k-1) = \sum_{(g',i',j) \in \mc{I}_{g,i}(k-1)} V_{g',i'}(j),
\end{equation}
where $\mc{I}_{g,i}(k-1)$ is the set containing all the indices $(g',i',j)$ for which the source $(g', i')$ is scheduled to transmit its $j$-th update during the $(k-1)$-th update to the beginning of the $k$-th update of source $(g,i)$.
\begin{figure}
	\center{\includegraphics[width=6in]{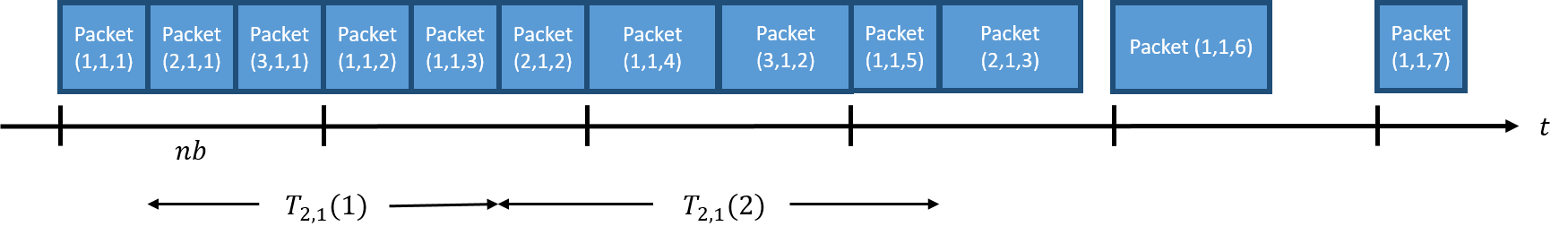}}
   	\caption{An example of packet transmission under GRR policy. In this example, we have $\mc{I}_{2,1}(1) = \{(2,1,1), (3,1,1), (1,1,3)\}$ and $\mc{I}_{2,1}(2) = \{(2,1,2), (1,1,4), (3,1,2),(1,1,5)\}$}
   	\label{fig:illustration_example2}
\end{figure}

\begin{example}
Let us take Example~\ref{exp:three_group_system} with only one source in each group for example. An illustration of the total transmission time and corresponding index set for this example can be found in Fig. \ref{fig:illustration_example2}. For the source 1 in group 2, one observes that

\begin{align}
    T_{2,1}(1) &= V_{2,1}(1) + V_{3,1}(1) + V_{1,1}(2) + V_{1,1}(3), \\
    T_{2,1}(2) &= V_{2,1}(2)+V_{1,1}(4)+V_{3,1}(2)+V_{1,1}(5), \label{eqn:T_2(3)} \\
    T_{2,1}(3) &= V_{2,1}(3)+V_{1,1}(6)+V_{1,1}(7).
\end{align}
Since the transmission time is i.i.d., both $T_{2,1}(1)$ and $T_{2,1}(2)$ are composed of four i.i.d transmission times and $T_{2,1}(3)$ is composed of three i.i.d transmission times, they are statistically not the same, highlighting the difference on $T_{g,i}(k-1)$ for different $k$.
\end{example}

We are now ready to study the expression of peak AoI. The departure time $D_{g,i}(k)$ of the packet $(g,i,k)$ can be described by
\begin{equation}\label{eqn:D_i(k)}
    D_{g,i}(k) = S_{g,i}(k) + W_{g,i}(k) + V_{g,i}(k),
\end{equation}
where $W_{g,i}(k)$ is the waiting time of the packet $(g,i,k)$. Now, plugging \eqref{eqn:D_i(k)} into \eqref{eqn:A(k)} leads to
\begin{align}
    A_{g,i}(k)
    &= S_{g,i}(k) + W_{g,i}(k) + V_{g,i}(k) - S_{g,i}(k-1) \label{eqn:A(k) S+W+V}\\
    &= W_{g,i}(k) + V_{g,i}(k) + d_g nb. \label{eqn:A(k) FCFS}
\end{align}
Hence, to analyze AoI, it suffices to trace the evolution of the waiting time, which we provide a recursive formula in the sequel.
\begin{lemma}\label{lemma:W(k) expressed by W(k-1)}
	The waiting time of the packet $(1,1,k)$ can be expressed recursively as follows,
\begin{align}\label{eqn:W(k) FCFS}
			W_{1,1}(k) = &\left(W_{1,1}(k-1) + V(k-1) - nb \right)^+,
		\end{align}
  where $V(k-1)=T_{1,1}(k-1)$ is the total transmission time between the $(k-1)$-th and $k$-th updates of source $(1,1)$.
\end{lemma}
\begin{IEEEproof}
Since the packet $(1,1,k)$ can be served after all scheduled packets in round $k-1$ are served, $W_{1,1}(k)$ depends on $D_{g^*,n_{g^*}}\left(k_{g^*}\right)$ where we use the notations $g^*$ , $n_{g^*}$, and $k_{g^*}$ to denote that the last updated packet before $(1,1,k)$ is the $k_{g^*}$-th update of source $(g^*,n_{g^*})$. If $D_{g^*,n_{g^*}}(k_{g^*}) > S_{1,1}(k)$, the BS will serve the packet $(1,1,k)$ immediately after the successful departure of the packet $(g^*,n_{g^*},k_{g^*})$. Thus, we have
    \begin{align}\label{eqn:W(k) express by W(k-1) in proof}
    W_{1,1}(k) &= D_{g^*,n_{g^*}}(k_{g^*}) - S_{1,1}(k) \nonumber\\
    &\overset{(a)}{=} S_{g^*,n_{g^*}}(k_{g^*}) + W_{g^*,n_{g^*}}(k_{g^*}) +V_{g^*,n_{g^*}}(k_{g^*}) - S_{1,1}(k) \nonumber\\
    &\overset{(b)}{=} W_{g^*,n_{g^*}}(k_{g^*}) + V_{g^*,n_{g^*}}(k_{g^*}) - nb \nonumber\\
    &\overset{(c)}{=} W_{1,1}(k-1) + \sum_{(g',i',j)\in \mc{I}_{1,1}(k) \setminus (g^*,n_{g^*},k_{g^*})} V_{g',i'}(j) +  V_{g^*,n_{g^*}}(k_{g^*}) - nb \nonumber\\    &\overset{(d)}{=} W_{1,1}(k-1) + V(k-1) - nb,
    \end{align}
    where (a) follows from \eqref{eqn:D_i(k)}, (b) is because of $S_{g^*,n_{g^*}}(k_{g^*})=S_{1,1}(k-1)$ caused by batch arrival. Recall that the packets arrive in batch at time $S_{1,1}(k-1)$ so that all the packets are stored in the queue waiting to be transmitted. (c) intends to express $W_{g^*,n_{g^*}}(k_{g^*})$ in terms of $W_{1,1}(k-1) + \sum_{(g',i',j)\in \mc{I}_{1,1}(k) \setminus (g^*,n_{g^*},k_{g^*})} V_{g',i'}(j)$, where the term $\sum_{(g',i',j)\in \mc{I}_{1,1}(k) \setminus (g^*,n_{g^*},k_{g^*})} V_{g',i'}(j)$ represents the overall transmission time of the packets scheduled between packets $(1,1,k-1)$ and $(g^*,n_{g^*},k_{g^*})$. (d) follows from the definition of $T_{1,1}(k-1)$ in \eqref{eqn:T_g(k)} and $V(k-1)=T_{1,1}(k-1)$. If $D_{g^*,n_{g^*}}(k_{g^*})\leq S_{1,1}(k)$, the BS is idle after transmitting the packet $(g^*,n_{g^*},k_{g^*})$ and we can immediately transmit the  packet (1,1,k) at the packet's arrival. Therefore, the waiting time is zero, which again fulfills \eqref{eqn:W(k) FCFS} as $W_{1,1}(k-1) + V(k-1) - nb = D_{g^*,n_{g^*}}(k_{g^*}) - S_{1,1}(k)\leq 0$.
\end{IEEEproof}
Next, we derive the peak age of the packet $(g,i,k)$ by applying Lemma~\ref{lemma:W(k) expressed by W(k-1)}.

\begin{lemma}\label{lemma:A(k) iterative formulation}
The peak age of the packet $(g,i,k)$ is
	\begin{align}\label{eqn:A(k) iterative formulation}
	A_{g,i}(k) &= \max_{1\leq \ell \leq \tilde{k}} \left\{ \sum_{r=\ell}^{\tilde{k}-1} V(r) - (\tilde{k}-\ell)nb\right\} + \sum_{(g',i',j)\in \mc{J}^-_{g,i}(k)}V_{g',i'}(j) + d_g nb,
	\end{align}
	where $\tilde{k} = d_g(k-1)+1$ and $\mc{J}^-_{g,i}(k)$ is the set containing all the indices $(g',i',j)$ for which the source $(g', i')$ is scheduled to transmit its $j$-th update between the $d_g(k-1)$-th update of source $(1,1)$ to the beginning of the $k$-th update of source $(g,i)$.
\end{lemma}
\begin{IEEEproof}
    See Appendix~\ref{apx:proof_peak_age_formulation_FCFS}.
\end{IEEEproof}

Intuitively, the AoI evolution above can be reasoned as follows. Let $\ell^*$ be the maximizer of the first term of \eqref{eqn:A(k) iterative formulation}. When $\ell^*=\tilde{k}$, the first term of \eqref{eqn:A(k) iterative formulation} becomes zero, corresponding to the case that all the scheduled packets in the rounds before the one containing the $k$-th update of source $(g,i)$ has been cleared, making the inter-arrival time and transmission time of all the sources preceding $(g,i)$ in the same round the sole contribution of $A_{g,i}(k)$. On the other hand, for $1\leq \ell^* \leq \tilde{k}-1$, the first term of \eqref{eqn:A(k) iterative formulation} corresponds to the extra time the $k$-th update of source $(g,i)$ has to wait caused by unfinished updates of previous rounds.

\subsection{Age violation probability analysis under the IPQ}\label{subsec:asymptotic analysis FCFS}
Based on the AoI evolution in Lemma~\ref{lemma:A(k) iterative formulation}, an upper and a lower bounds on the age violation probability are derived.
 We first present the upper bound in the following theorem.
\begin{theorem}\label{thm:upper_bound_FCFS}
For any $x>0$, $k>0$, $\theta>0$, and $c'>0$, if $\sum_{j=1}^{\eta}\frac{\tilde{d}}{d_j}\alpha_j\Lambda_v(\theta) - \tilde{d}\theta b < 0$, then the age violation probability of source $(g,i)$ under the IPQ, is upper bounded as follows,
	\begin{equation}\label{eqn:age violation prob UB}
	    \text{Pr}\left(A_{g,i}(k)\geq nx\right) \leq (\tilde{d} c'+1) \exp\left\{ -n \cdot \min_{0 \leq \ell' \leq k'} \gamma_{g,i}^{(U,\ell')}(x,k,n)\right\},
	\end{equation}
	where
	\begin{align}
	    \gamma_{g,i}^{(U,\ell')}(x,k,n)
        &=\begin{cases}
        I_{g,i}^{(U,0)}\left( x - d_g b,k,n \right), &\text{for } \ell'=0,\\
        \ell' I_{g,i}^{(U,\ell')}\left(\frac{x}{\ell'}+\frac{(\ell'-1) \tilde{d}-d_g}{\ell'}b,k,n\right), &\text{for } 1 \leq \ell' \leq k',
        \end{cases}\label{eqn:gamma_UB}
	\end{align}
and
\begin{align}
    &I_{g,i}^{(U,0)}(x,k,n) = \sup_{\theta} \left\{ \vphantom{\frac{C(\ell)  \sum_{j=1}^\eta \frac{\tilde{d}}{d_j}\alpha_j }{C(\ell)} \Lambda_v(\theta)}\theta x - m_{g,i}(k) \Lambda_v(\theta) \right\}, &\text{for } \ell'=0,\label{eqn:rate_function_ell=0 FCFS}\\
    &I_{g,i}^{(U,\ell')}(x,k,n) = \sup_{\theta} \left\{ \theta x - \sum_{j=1}^\eta \frac{\tilde{d}}{d_j}\alpha_j  \Lambda_v(\theta) -\frac{m_{g,i}(k)}{\ell'}\Lambda_v(\theta)\right\}, &\text{for } 1 \leq \ell' \leq k'.\label{eqn:rate_function_ell!=0 FCFS}
\end{align}
Here $\ell' = \lceil \frac{\Tilde{k}-\ell}{\Tilde{d}} \rceil$, $k'=\lceil \frac{\Tilde{k}-1}{\Tilde{d}} \rceil$, and $m_{g,i}(k)=|\mc{J}^-_{g,i}(k)|/n$ is the ratio of the number of transmitted packet in the set $\mc{J}^-_{g,i}(k)$ to the total number of sources $n$ in the system. 
\end{theorem}
\begin{IEEEproof}
    See Appendix~\ref{apx:proof_upper_bound_FCFS}.
\end{IEEEproof}
The lower bound is presented in the following theorem.
\begin{theorem}\label{thm:lower_bound_FCFS}
For any $x>0$, $k>0$, $\theta>0$, and $\epsilon>0$, let $\ell'=\lceil \frac{\Tilde{k}-\ell}{\Tilde{d}} \rceil$ and $k'=\lceil \frac{\Tilde{k}-1}{\Tilde{d}} \rceil$, the age violation probability of source $(g,i)$ under the IPQ is lower bounded as follows,
	\begin{equation}\label{eqn:age violation prob LB}
    \text{Pr}\left(A_{g,i}(k)\geq nx \right) \geq \exp\left\{ -n \left(  \min_{0 \leq \ell' \leq k'} \gamma_{g,i}^{(L,\ell')}(x,k,n) + \epsilon \right)\right\},
\end{equation}
where
\begin{equation}
    \gamma_{g,i}^{(L,\ell')}(x,k,n) =
    \begin{cases}
    I_{g,i}^{(L,0)}\left( x-d_g b,k,n\right), &\text{for } \ell'=0\\
        \ell' I_{g,i}^{(L,\ell)} \left( \frac{x}{\ell'}+\frac{\ell'   \Tilde{d} - d_g}{\ell'}b,k,n\right),
    &\text{for } 1 \leq \ell' \leq k',
    \end{cases}
\end{equation}
and
\begin{align}
    &I_{g,i}^{(L,0)}(x,k,n) = \sup_{\theta} \left\{ \theta x - m_{g,i}(k) \Lambda_v(\theta) \right\}, &\text{for } \ell'=0,\\
    &I_{g,i}^{(L,\ell')}(x,k,n) = \sup_{\theta} \left\{ \theta x - \frac{\left( \ell'-1 \right) \sum_{j=1}^{\eta} \frac{\Tilde{d}}{d_j}\alpha_j }{\ell'}   \Lambda_v(\theta) -\frac{m_{g,i}(k)}{\ell'}\Lambda_v(\theta)\right\}, &\text{for } 1 \leq \ell' \leq k'.
\end{align}
\end{theorem}
\begin{IEEEproof}
    See Appendix~\ref{apx:proof_lower_bound_FCFS}.
\end{IEEEproof}
To characterize the asymptotic behavior of the proposed GRR scheduling policy in the large system regime, the following two corollaries are derived from the above theorems.

\begin{corollary}\label{coro:asym_UB_FCFS}
For any $x>0$, $k>0$, and $\ell'$ and $k'$ as defined in Theorem~\ref{thm:upper_bound_FCFS}, the asymptotic decay rate of source $(g,i)$ under the IPQ is upper and lower bounded by
\begin{align}\label{eqn:asym_LB_FCFS}
    -\lim_{n \to \infty} \frac{1}{n} \log \text{Pr}\left(A_{g,i}(k)\geq nx\right) \leq \min_{0\leq \ell' \leq k'} \gamma_{g,i}^{(L,\ell')}(x,k,\infty),
\end{align}
and
\begin{align}\label{eqn:asym_UB_FCFS}
    -\lim_{n \to \infty} \frac{1}{n} \log \text{Pr}\left(A_{g,i}(k)\geq nx\right) \geq \min_{0\leq \ell' \leq k'} \gamma_{g,i}^{(U,\ell')}(x,k,\infty),
    \end{align}
respectively.
\end{corollary}

So far, we have focused on the age violation probability and its asymptotic decay rate for a particular update $k$. Next, we want to analyze the long-run fraction of  violation probability of source $(g,i)$
    $     \lim_{\kappa\to \infty} \frac{1}{\kappa} \sum_{k=1}^\kappa \mathbb{E} \left[ \mathbbm{1}_{\{A_{g,i}(k)\geq nx\}}\right].$
    Re-grouping the summation according to $\zeta=k\mmod \tilde{d}/d_g$ and apply the upper bound of violation probability, we have
    \begin{corollary}\label{coro:long-run fraction of violation probability}
    The long-run fraction of age violation probability of source $(g,i)$ is upper bounded as follow:
    \begin{align}\label{eqn:long_run_UB}
    \lim_{\kappa\to \infty} \frac{1}{\kappa} \sum_{k=1}^\kappa \mathbb{E} \left[ \mathbbm{1}_{\{A_{g,i}(k)\geq nx\}}\right] \leq \sum_{\zeta=1}^{\Tilde{d}/d_g} \frac{d_g}{\Tilde{d}} (\Tilde{d}c'+1) e^{-n \min_{\ell' \geq 0}\gamma_{g,i}^{(U,\ell')}(x,\zeta,\infty)}.
    \end{align}
    Moreover, the asymptotic decay rate of the long-run fraction of age violation probability for source $(g,i)$ is given by
    \begin{align}
    -\lim_{n\to \infty} \frac{1}{n} \log \lim_{\kappa\to \infty} \frac{1}{\kappa} \sum_{k=1}^\kappa \mathbb{E} \left[ \mathbbm{1}_{\{A_{g,i}(k)\geq nx\}} \right] \geq \sum_{\zeta=1}^{\Tilde{d}/d_g} \min_{\ell'\geq 0} \gamma_{g,i}^{(U,\ell')}(x,\zeta,\infty),\label{eqn:long_run_asymptotic_UB}
\end{align}
    where $\gamma_{g,i}^{(U,\ell')}\left( x,\zeta,\infty \right)$ is defined in \eqref{eqn:gamma_UB}.
    \end{corollary}
    \begin{IEEEproof}
        The proof of \eqref{eqn:long_run_UB} is relegated to Appendix \ref{apx:proof_of_long-run probability}. To prove \eqref{eqn:long_run_asymptotic_UB}, we note that from \eqref{eqn:long_run_UB}
        \begin{align}
    &-\lim_{n\to \infty} \frac{1}{n} \log \lim_{\kappa\to \infty} \frac{1}{\kappa} \sum_{k=1}^\kappa \mathbb{E} \left[ \mathbbm{1}_{\{A_{g,i}(k)\geq nx\}} \right]
    \geq -\lim_{n\to \infty} \frac{1}{n} \log \left( \sum_{\zeta=1}^{\Tilde{d}/d_g} \frac{d_g}{\Tilde{d}} (\Tilde{d}c'+1) e^{-n \min_{\ell' \geq 0}\gamma_{g,i}^{(U,\ell')}(x,\zeta,\infty)} \right) \nonumber \\
    &\overset{(a)}{\geq} \lim_{n\to \infty} \frac{1}{n} \left[ -\log \left( \frac{d_g}{\Tilde{d}} (\Tilde{d}c'+1) \right) +n \sum_{\zeta=1}^{\Tilde{d}/d_g} \min_{\ell'\geq 0} \gamma_{g,i}^{(U,\ell')}(x,\zeta,\infty)  \right] = \sum_{\zeta=1}^{\Tilde{d}/d_g} \min_{\ell'\geq 0} \gamma_{g,i}^{(U,\ell')}(x,\zeta,\infty).
\end{align}
where (a) follows from the Jensen's inequality.
    \end{IEEEproof}

\subsection{Approximation of the asymptotic decay rate under the IPQ}\label{subsec:aprrox_FCFS}
To gain further insight, we specialize our results to a specific distribution for $V_{g,i}(k)\sim \mathsf{Exp}(\lambda)$ whose optimal $\theta$ can be solved in closed-form. This choice of distribution is not merely for mathematical convenience but also for its connection to transmission over packet erasure channel (PEC), which will be elaborated in Section~\ref{sec:numerical and simulation}.
Moreover, we consider $x > d_gb$; otherwise, the age violation probability would be 1 and the discussion would become meaningless.

Note that it is not difficult to see that the rate functions are convex in $\theta$. Thus, we differentiate the rate functions in \eqref{eqn:rate_function_ell=0 FCFS} and \eqref{eqn:rate_function_ell!=0 FCFS} to obtain
\begin{align}
    &\theta_{g,i}^{*(U,0)}(x,k,\infty) = \lambda - \frac{m_{g,i}(k)}{x},  \label{eqn:theta_star ell=0}
\end{align}
and
\begin{align}
    &\theta_{g,i}^{*(U,\ell)}(x,k,\infty) = \lambda - \frac{\ell' \sum_{j=1}^\eta \frac{\Tilde{d}}{d_g}\alpha_j+m_{g,i}(k)}{x} \label{eqn:theta_star 1<ell<k},
\end{align}
respectively. The $\theta_{g,i}^{*(U,\ell)}(x,k,\infty)$ is then plugged into $\gamma_{g,i}^{(U,\ell)}(x,k,\infty)$, leading to
\begin{align}
    \gamma_{g,i}^{(U,0)}(x,k,\infty) 
    &= \lambda x -d_g \left( \lambda b - \frac{m_{g,i}(k)}{x} b \right) -m_{g,i}(k) - m_{g,i}(k) \log \left( \frac{\lambda x}{m_{g,i}(k)} \right) ,\label{eqn:gamma ell=0 theta*}
\end{align}
and
\begin{align}
    \gamma_{g,i}^{(U,\ell)}&(x,k,\infty)
    = \lambda x -d_g \left[ \lambda b - \left(\ell'\sum_{j=1}^\eta \frac{\Tilde{d}}{d_j}\alpha_j + m_{g,i}(k) \right) \frac{b}{x} \right] - \tilde{d}(\ell'-1)\left( \ell' \sum_{j=1}^\eta \frac{\tilde{d}}{d_j} \alpha_j + m_{g,i}(k) \right)  \frac{b}{x}\nonumber\\
    &- \left( \ell' \sum_{j=1}^\eta \frac{\Tilde{d}}{d_j} \alpha_j + m_{g,i}(k)\right) \left[ 1+ \log\left( \frac{\lambda x}{\ell' \sum_{j=1}^\eta \frac{\Tilde{d}}{d_j}\alpha_j+m_{g,i}(k)} \right) \right] + (\ell'-1) \Tilde{d} \lambda b\label{eqn:gamma ell!=0 theta*}.
\end{align}

In what follows, we separately discuss the large transmission rate and the small transmission rate cases.

\underline{Case 1 (Large transmission rate):}\\
In this case, we assume $b \gg \frac{1}{\lambda}$. Under this assumption, the packets stored in queues are very likely to be cleared before the arrival of the next batch. From \eqref{eqn:gamma ell=0 theta*} and \eqref{eqn:gamma ell!=0 theta*}, using the assumption $b \gg \frac{1}{\lambda}$, we have
\begin{align}
    \gamma_{g,i}^{(U,0)}(x,k,\infty) \approx \lambda ( x-d_g b),\label{eqn:approx_gamma ell=0 theta* large transmission rate}
\end{align}
and
\begin{align}
    \gamma_{g,i}^{(U,\ell)}(x,k,\infty) \approx &\lambda x -d_g \lambda b - \tilde{d}(\ell'-1)\left( \ell' \sum_{j=1}^\eta \frac{\tilde{d}}{d_j} \alpha_j + m_{g,i}(k) \right)  \frac{b}{x}  + (\ell'-1) \Tilde{d} \nonumber\\
    &- \left( \ell' \sum_{j=1}^\eta \frac{\Tilde{d}}{d_j} \alpha_j + m_{g,i}(k) \right) \left[ 1 + \log\left( \frac{\lambda x}{\ell' \sum_{j=1}^\eta \frac{\Tilde{d}}{d_j}\alpha_j + m_{g,i}(k) }\right) \right], \label{eqn:approx_gamma ell!=0 theta* large transmission rate}
\end{align}
respectively.
The approximations in \eqref{eqn:approx_gamma ell=0 theta* large transmission rate} and \eqref{eqn:approx_gamma ell!=0 theta* large transmission rate} allow us to draw the conclusion for this case that the exponents of the age violation probability are larger for sources with smaller $d_g$.

\underline{Case 2 (Small transmission rate):}\\
We assume $b \approx \frac{1}{\lambda}$ and $b-\frac{1}{\lambda}>\epsilon$ where $\epsilon>0$ is a small value to ensure that the queue is stable and we are in the large buffer regime. In this case, every packet has to wait in the queue for a long period of time with high probability. We then approximate \eqref{eqn:gamma ell=0 theta*} and \eqref{eqn:gamma ell!=0 theta*} to get
\begin{align}
    \gamma_{g,i}^{(U,0)}(x,k,\infty) &\approx \lambda x - d_g \left(1-\frac{m_{g,i}(k)}{x}b \right) - m_{g,i}(k) - m_{g,i}(k) \log \left( \frac{\lambda x}{m_{g,i}(k)} \right) \nonumber \\
    &\leq \lambda x -d_g \left( 1-\frac{m_{g,i}(k)}{x}b \right),\label{eqn:approx_gamma ell=0 theta* small transmission rate}
\end{align}
and
\begin{align}\label{eqn:approx_gamma ell!=0 theta* small transmission rate}
    \gamma_{g,i}^{(U,\ell)}(x,k,\infty) \approx &\lambda x -d_g \left[ 1-\left( \ell' \sum_{j=1}^\eta \frac{\Tilde{d}}{d_j} \alpha_j + m_{g,i}(k)\right) \frac{b}{x} \right]   - \tilde{d}(\ell'-1)\left( \ell' \sum_{j=1}^\eta \frac{\tilde{d}}{d_j} \alpha_j + m_{g,i}(k) \right)  \frac{b}{x} \nonumber\\
    & - \left( \ell' \sum_{j=1}^\eta \frac{\Tilde{d}}{d_j} \alpha_j + m_{g,i}(k) \right) \left[ 1+ \log \left( \frac{\lambda x}{\ell' \sum_{j=1}^\eta \frac{\Tilde{d}}{d_j}\alpha_j}+ m_{g,i}(k)\right) \right],
\end{align}
respectively.
From \eqref{eqn:approx_gamma ell=0 theta* small transmission rate} and \eqref{eqn:approx_gamma ell!=0 theta* small transmission rate}, same conclusion can be drawn that the age violation probability decays faster for the source with smaller $d_g$.

\subsection{Homogeneous case under the IPQ}\label{subsec:RR_FCFS}
We now specialize our results to the homogeneous system in which all sources' arrival periods are the same. That is, $d_g = 1$ for all $g \in [G]$. i.e., all the sources can be regarded as in the same group. We thus drop the subscript $g$ in all random variables in this subsection.

The waiting time in homogeneous system can be directly obtained from Lemma \ref{lemma:W(k) expressed by W(k-1)},
\begin{equation}\label{eqn:W(k) RR FCFS}
    W_1(k) = \left( W_1(k-1) + V(k-1) - nb \right)^+,
\end{equation}
where $V(k-1)=\sum_{i=1}^n V_{i}(k-1)$ in this case. Iteratively plugging \eqref{eqn:W(k) RR FCFS} into \eqref{eqn:A(k) FCFS} shows the expression of the peak age for the $k$-th updated packet of source $i$ as
\begin{align}
    A_i(k) &= \max_{1\leq \ell \leq k} \left\{ \sum_{r=\ell}^{k-1} V(r) - (k-\ell)nb\right\} + \sum_{(i,j)\in \mc{J}^-_{i}(k)}V_{i}(j) + nb, \label{eqn:A(k) RR FCFS}
\end{align}
coinciding with the peak age can be deduced from Lemma \ref{lemma:A(k) iterative formulation}.
Upper and lower bounds on the age violation probability for the homogeneous case can then be obtained from Theorems \ref{thm:upper_bound_FCFS} and \ref{thm:lower_bound_FCFS}, which are summarized in the following corollary.
\begin{corollary}
For any $x>0$, $k>0$, and $\ell$ as defined in Theorem \ref{thm:upper_bound_FCFS}, the age violation probability of source $i$ under the IPQ in the homogeneous case is upper and lower bounded by
\begin{align}
    &\text{Pr}\left(A_i(k)\geq nx\right) \leq c \cdot \exp \left[ -n \min_{0 \leq \ell' \leq k-1} \gamma_{i}^{(\ell')}(x,k,n) \right]  \label{eqn:upper bound_bound RR FCFS},
\end{align}
and
\begin{align}
    &\text{Pr}\left(A_i(k)\geq nx\right) \geq \exp \left[ -n\left( \min_{0 \leq \ell' \leq k-1} \gamma_i^{(\ell')}(x,k,n) + \epsilon\right) \right], \label{eqn:lower_bound RR FCFS}
\end{align}
respectively, where $c>0$ is a constant, $\epsilon>0$ can be arbitrarily small, and
\begin{align}
	&\gamma_{i}^{(\ell')}(x,k,n)
        =\begin{cases}
        \hat{I}_{i}^{(0)}\left( x - b,k,n \right), \quad \text{for } \ell'=0,\\
        \ell' \hat{I}_{i}^{(\ell')} \left(\frac{x}{\ell'}+\frac{(\ell'-1)}{\ell'}b,k,n\right), \quad \text{for } 1 \leq \ell' \leq k',
        \end{cases}
\end{align}
with
    $\hat{I}_i^{(\ell')}(x,k,n) = \sup_{\theta:\Lambda_v(\theta) - \theta b < 0} \left[ \theta x - \frac{\ell'+ i/n}{\ell'} \Lambda_v(\theta) \right].$
\end{corollary}

The asymptotic scaling of the age violation probability for the homogeneous case can be exactly characterized as follows.
\begin{corollary}\label{coro:asym_FCFS RR}
For any $x>0$, the asymptotic decay rate of source $i$ under the IPQ in homogeneous case is given by
\begin{equation}\label{eqn:asym_FCFS RR}
    -\lim_{n \to \infty} \frac{1}{n} \log \text{Pr}(A_i(k)\geq nx) = \min_{0 \leq \ell' \leq k-1} \gamma_{i}^{(\ell')}(x,k,\infty).
\end{equation}
\end{corollary}

\section{Proposed GRR with preemption under the SPQ}\label{sec:LCFS}
In this section, we turn our attention to the SPQ case. We first propose the GRR scheduling policy for SPQ in Section~\ref{subsec:GRR single packet queue} and analyze the evolution of AoI under GRR in Section~\ref{subsec:age analysis preemptive}. An upper bound on the age violation probability are derived in Section~\ref{subsec:asymptotic analysis preemptive} followed by its approximation in Section~\ref{subsec:aprrox_LCFS}. The homogeneous case is then consider in Section~\ref{subsec:RR_LCFS}.

\subsection{Proposed GRR under the SPQ}\label{subsec:GRR single packet queue}
We employ the same GRR scheduling policy in Definition~\ref{def:GRR} to decide the schedule among sources. However,  unlike the the infinite packet queueing discipline, we store the latest packet only and preempt all the old packets for each source. Intuitively, this prevents the BS from updating stale information, resulting in a smaller AoI as compared to the infinite packet queueing discipline. Our analysis in Section~\ref{subsec:age analysis preemptive} makes this intuition precise.

\subsection{Age analysis under the SPQ}\label{subsec:age analysis preemptive}

We first recall that the peak age of the packet $(g,i,k)$ is defined in \eqref{eqn:A(k)} and the departure time is given in \eqref{eqn:D_i(k)}. Plugging \eqref{eqn:D_i(k)} into \eqref{eqn:A(k)} shows the age expression in \eqref{eqn:A(k) S+W+V}, making the derivation of waiting time the main challenge. To proceed, we define two new random variables $N_{g,i}(k-1)$ and $p_{g,i}(k-1)$, where $N_{g,i}(k-1)$ is the idle time occurring after serving the $(k-1)$-th update from source $(g,i)$ and $p_{g,i}(k-1)$
is the number of source $i$'s packets preempted between the $(k-1)$-th and the $k$-th updates.

An example with three sources in one group is illustrated in Fig.~\ref{fig:LCFS_ex} in which the variables $N_{g,i}(k-1)$ and $p_{g,i}(k-1)$ are explained.
In Fig.~\ref{fig:LCFS_case1}, the first round has a fairly long transmission time of $V_{1,1}(1)+V_{1,2}(1)+V_{1,3}(1)>2nb$; hence, $N_{1,1}(1)=0$ and $p_{1,1}(1)=1$. On the contrary, the second round has a short overall transmission time; hence, the BS has to wait $N_{1,1}(2)$ until the arrival of the next batch to start the service of source 1 in the third round. In this case, $p_{1,1}(2)=0$. Another realization is shown in
Fig.~\ref{fig:LCFS_case2}. In this figure, we see that $V_{1,1}(1)$ is fairly large, which results in the preemption of one packet from the source 2 and 3. After $V_{1,2}(1)$, $V_{1,3}(1)$, and $V_{1,1}(2)$, the packets arrived at $2nb$ are all served and the BS again becomes idle. This shows that the idle time $N_{1,i}(k-1)$ does not necessarily occur after serving an entire round. Fig.~\ref{fig:LCFS_case3} shows another case that the BS becomes idle after serving a packet from source 2. 

With the introduction of the two new random variables, we obtain the expression of waiting time under the SPQ in the following lemma:
\begin{figure}
	\centering
    \begin{subfigure}[b]{2.8in}
	\includegraphics[width=\textwidth]{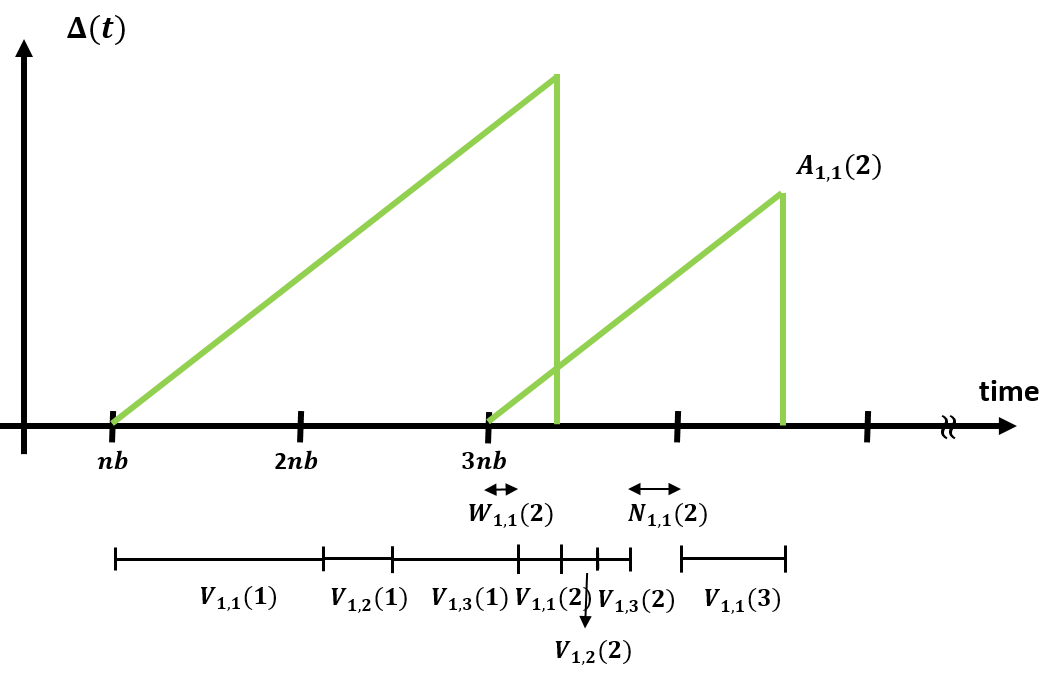}
   	\caption{\label{fig:LCFS_case1} Idle time $N_{1,1}(2)$ occurs after serving the third source.}
    \end{subfigure}
    \hfill
    \begin{subfigure}[b]{2.8in}
	\includegraphics[width=\textwidth]{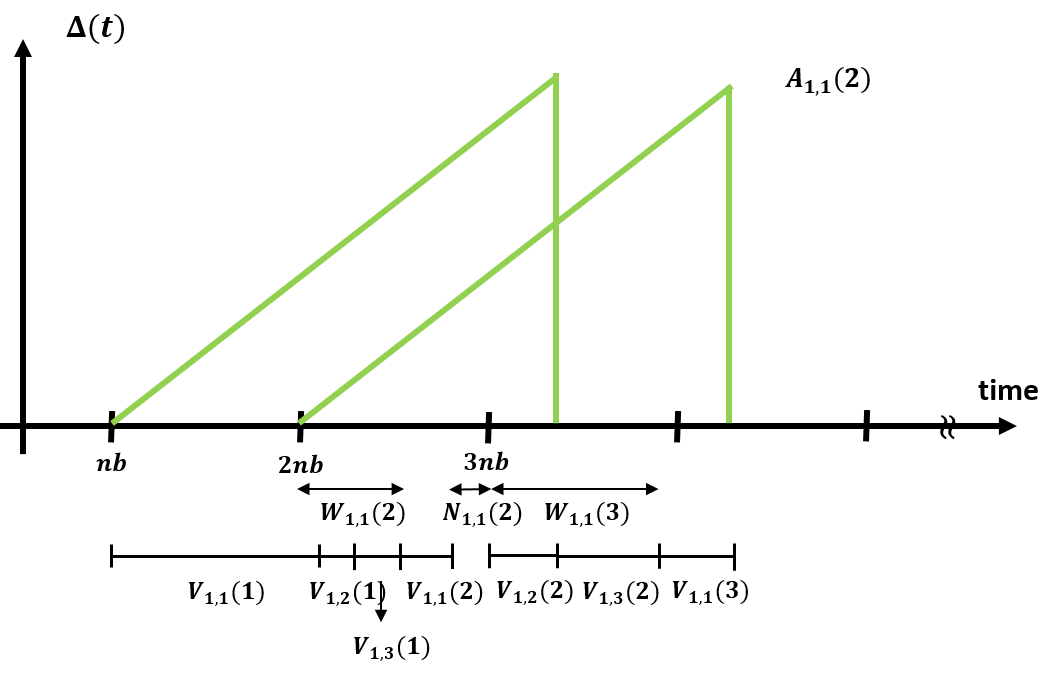}
   	\caption{\label{fig:LCFS_case2} Idle time $N_{1,1}(2)$ occurs after serving the first source.}
    \end{subfigure}
    \hfill
    \begin{subfigure}[b]{2.8in}
	\includegraphics[width=\textwidth]{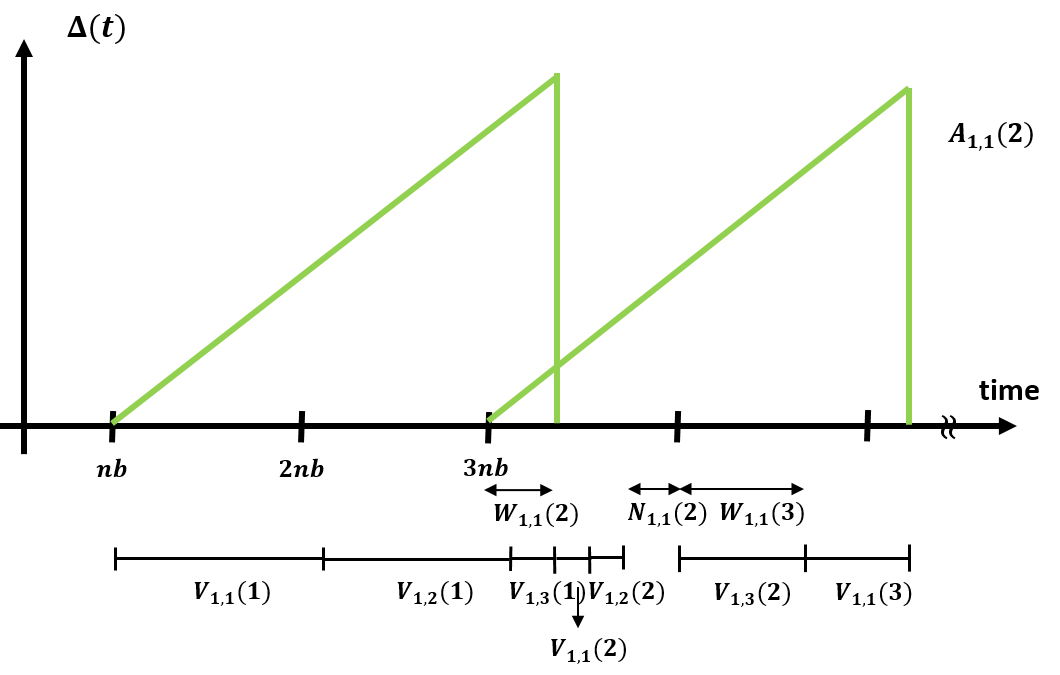}
   	\caption{\label{fig:LCFS_case3} Idle time $N_{1,1}(2)$ occurs after serving the second source.}
    \end{subfigure}
    \caption{An example of AoI under the SPQ.}
    \label{fig:LCFS_ex}
    \vspace{-20pt}
\end{figure}
\begin{lemma}\label{lemma:waiting_time_LCFS}
    The waiting time of the packet $(g,i,k)$ can be expressed as follows
    \begin{align}\label{eqn:W(k) LCFS}
        W_{g,i}(k) &= W_{g,i}(k-1) + T_{g,i}(k-1) + N_{g,i}(k-1) - \left( p_{g,i}(k-1)+1 \right)  d_g nb.
    \end{align}
    \end{lemma}
    \begin{IEEEproof}
        We first note that without preemption, the waiting time of the $k$-th updated packet from source $i$ would be the sum of $W_{g,i}(k-1)$ the waiting time in the previous round, $T_{g,i}(k-1)$ the total transmission time since transmitting the packet $(g,i,k-1)$ to the beginning of the transmission of packet $(g,i,k)$, and $N_{g,i}(k-1)$ the idle time to fill the gap (if exists) in the procedure of $T_{g,i}(k-1)$. i.e., it would be
        \begin{equation}\label{eqn:W_i(k) by W_i(k-1) proof}
            W_{g,i}(k-1) + T_{g,i}(k-1) + N_{g,i}(k-1) - d_g nb.
        \end{equation}
        Now, if $\eqref{eqn:W_i(k) by W_i(k-1) proof}\geq d_g nb$, the $p_{g,i}(k-1)$ stale packets will be preempted, which leads us to \eqref{eqn:W(k) LCFS} and completes the proof.

    \end{IEEEproof}


Next, we provide the formulation of the peak age of the packet $(g,i,k)$.

\begin{lemma}\label{lemma:peak_age_formulation_preemption}
		The peak age of the packet $(g,i,k)$ under the SPQ is
	\begin{align}
		A_{g,i}(k) = \max &\left\{ W_{g,i}(k-1) + \sum_{g',i',j \in \mc{J}^+_{g,i}(k-1)} V_{g',i'}(j) + \sum_{r=d_g(k-2)+2}^{d_g(k-1)} V(r), \right. \nonumber\\
        &\left. \max_{2 \leq \ell \leq d_g+1} \left[ \sum_{r=d_g(k-2)+2}^{d_g(k-1)} V(r) + (\ell-1) nb \right] \right\} + \sum_{(g',i',j) \in \mc{J}^-_{g,i}(k)} V_{g',i'}(j).\label{eqn:peak_age_preemptive}
		\end{align}
\end{lemma}
where $\mc{J}^+_{g,i}(k)$ contains all the indexes $(g',i',j)$ such that transmitting the packet $(g,i,k)$ to the end of that round.
\begin{IEEEproof}
    See Appendix \ref{apx:proof_peak_age_formulation_LCFS}.
\end{IEEEproof}

\subsection{Age violation probability analysis under the SPQ}\label{subsec:asymptotic analysis preemptive}
With the evolution of AoI above, we are now ready to present an upper bound on the age violation probability of peak age. 

\begin{theorem}\label{thm:upper_bound_preemptive}
The age violation probability of the packet $(g,i,k)$ under the SPQ is upper bounded by
		\begin{equation}\label{eqn:age violation prob UB preemptive}
			\text{Pr}( A_{g,i}(k)\geq nx) \leq \exp \left\{ -n \tilde{I}_{g,i} \left( x- d_g b ,k,n\right) \right\},
		\end{equation}
where
\begin{equation}\label{eqn:rate function preemptive}
    \tilde{I}_{g,i}(x,k,n) = \sup_{\theta}\left\{ \theta x - \left( t_{g,i}(k-1)+\frac{1}{n} \right) \Lambda_v(\theta) \right\},
\end{equation}
with $t_{g,i}(k-1)=|\mathcal{I}_{g,i}(k-1)|/n$ being the ratio of the number of transmissions between transmitting the packet $(g,i,k-1)$ and the packet $(g,i,k)$ to the total number of sources $n$ in the system.
\end{theorem}
\begin{IEEEproof}
	The proof is relegated to Appendix~\ref{apx:proof_upper_bound_preemptive}.
\end{IEEEproof}
\begin{remark}
    With \eqref{eqn:peak_age_preemptive}, it is possible to use techniques similar to that in Section~\ref{subsec:asymptotic analysis FCFS} to obtain lower bounds. However, we are unable to find a good lower bound on $W_{g,i}(k-1)$ that result in a tight lower bound. We leave it as a future work.
\end{remark}

Two corollaries are in place whose proofs are omitted:


\begin{corollary}\label{coro:asym_preemptive}
Assume $\Lambda_v(\theta)$ is bounded. The asymptotic decay rate of the packet $(g,i,k)$ under the SPQ is bounded by
    \begin{equation}
        -\lim_{n \to \infty}\frac{1}{n} \log \text{Pr}(A_{g,i}(k)\geq nx) \geq \tilde{I}_{g,i}(x-d_g b, k, \infty), \label{eqn:asym_upper_bound_preemptive}
    \end{equation}
where
\begin{equation}\label{eqn:rate_function_single_packet_GRR_infty}
    \tilde{I}_{g,i}(x,k,\infty) =  \sup_{\theta} \left\{ \theta x - t_{g,i}(k-1)\Lambda_v(\theta) \right\}.
\end{equation}
\end{corollary}
\begin{corollary}
The asymptotic decay rate of source $(g,i)$ under the SPQ is bounded as follows,
\begin{align}
    &-\lim_{n\to \infty} \frac{1}{n} \log \lim_{\kappa\to \infty} \frac{1}{\kappa} \sum_{k=1}^\kappa \mathbbm{1}_{\{A_{g,i}(k)\geq nx\}} \geq \sum_{\zeta=1}^{\tilde{d}/d_g} \tilde{I}_{g,i} \left( x- d_g b ,\zeta,\infty \right),\label{eqn:upper_bound_preemptive}
\end{align}
\end{corollary}

\subsection{Approximation of the asymptotic decay rate under the SPQ}\label{subsec:aprrox_LCFS}
To gain insights, we again specialize our results to the case $V_{g,i}(k)\sim \mathsf{Exp}(\lambda)$. Differentiating the rate function in \eqref{eqn:rate_function_single_packet_GRR_infty} results in
\begin{equation}
    \theta^*_{g,i}(x,k,\infty) = \lambda - \frac{t_{g,i}(k-1)}{x}.
\end{equation}
Then plugging this $\theta^*_{g,i}(x,k,\infty)$ into \eqref{eqn:asym_upper_bound_preemptive} leads to
\begin{equation}
    \Tilde{I}_{g,i}(x-d_gb,k,\infty) = \lambda x -d_g \left( \lambda b - \frac{t_{g,i}(k-1)}{x}b \right) - t_{g,i}(k-1) - t_{g,i}(k-1) \log \left( \frac{\lambda x}{t_{g,i}(k-1)}\right).\label{eqn:asym_prob_theta_star_LCFS}
\end{equation}
In what follows, the large transmission rate and the small transmission rate cases are discussed.

\underline{Case 1 (Large transmission rate):}\\
In this case, we assume $b \gg \frac{1}{\lambda}$. \eqref{eqn:asym_prob_theta_star_LCFS} then leads to
\begin{equation}
    \Tilde{I}_{g,i}(x-d_gb,k,\infty) \approx \lambda x -  d_g \lambda b - t_{g,i}(k-1) - t_{g,i}(k-1) \log \left(\frac{\lambda x}{t_{g,i}(k-1)}\right) .
\end{equation}
The approximation indicates that for this case the age violation probability decays faster for sources with smaller $d_g$. 

\underline{Case 2 (Small transmission rate):}\\
We assume $b \approx \frac{1}{\lambda}$ and $b-\frac{1}{\lambda} > \epsilon$ where $\epsilon>0$ is a small value. We have
\begin{equation}
    \Tilde{I}_{g,i}(x-d_gb,k,\infty) \approx \lambda x- d_g \left(1- \frac{t_{g,i}(k-1)}{x}b \right) - t_{g,i}(k-1) - t_{g,i}(k-1) \log \left(\frac{\lambda x}{t_{g,i}(k-1)}\right),
\end{equation}
which allows us to draw the same conclusion as case 1.

\subsection{Homogeneous case under the SPQ}\label{subsec:RR_LCFS}
Here, we again focus on the homogeneous case, where the GRR policy reduces to the RR policy in Definition~\ref{def:RR} and we drop the first subscript $g$ for all random variables in this subsection. 
Applying $d_g=1$ to \eqref{eqn:peak_age_preemptive} results in
\begin{equation}\label{eqn:A(k) RR LCFS}
    A_i(k) = \max \left\{ W_i(k-1) + \sum_{i'=i}^n V_{i'}(k-1), nb \right\} + \sum_{i'=1}^i V_{i'}(k).
\end{equation}
In what follows, a bound on the age violation probability is derived. The proof of these results closely follows the steps in Section~\ref{subsec:asymptotic analysis preemptive} and is thus omitted. 
\begin{corollary}\label{coro:upper_bound_preemptive RR}
    The age violation probability of source $i$ under the SPQ in homogeneous system is bounded by
\begin{equation}\label{eqn:upper_bound_preemptive RR}
    \text{Pr}\left(A_i(k)\geq nx\right) \leq \exp \left\{ -n \Bar{I_i} \left( x- b,k,n\right) \right\},
\end{equation}
where
$    \Bar{I}_i(x,k,n) = \sup_{\theta}\left\{ \theta x - \left( \frac{n+1}{n} \right) \Lambda_v(\theta) \right\}.$
Moreover, the asymptotic decay rate of source $i$ under the SPQ in homogeneous system is  bounded by
\begin{equation}\label{eqn:asym_prob_single_packet_RR}
    -\lim_{n \to \infty}\frac{1}{n} \log \text{Pr}\left( A_i(k)\geq nx\right) \geq  \Bar{I}_i(x-b,k,\infty),
\end{equation}
where
\begin{equation}\label{eqn:rate_function_single_packet_infty}
    \Bar{I}_i(x,k,\infty) = \sup_{\theta>0} \left\{ \theta x - \Lambda_v(\theta) \right\}.
\end{equation}
\end{corollary}

\section{Discussion and Simulation Results}\label{sec:numerical and simulation}
In this section, we use computer simulation to verify our analysis. We assume that the system consists of three heterogeneous groups of sources. Each group has the same number of sources $n$. The arrival period of group $g\in\{1,2,3\}$ scales linearly with $n$, given by $d_g nb$ with $b=5$ and $(d_1,d_2,d_3)=(1,2,4)$. Each transmission time follows the exponential distribution with parameter $\lambda$, i.e., $\mathsf{Exp}(\lambda)$. Here, $\lambda $ can be understood as the service rate, which makes $1/\lambda$ the average transmission time. We also consider a more practical setting in which time is discretized into slots and each scheduled packet is transmitted at the begining of a slot to the destination through a PEC with erasure probability $\varepsilon=0.7165$. If the packet is successfully received at the destination, an ACK signal is fedback at the end of the slot. Otherwise, the BS retransmits the packet in the next slot. This gives rise to the geometrically distributed transmission time with successful probability $1-\varepsilon=0.2835$. It is quite well-known that exponential distribution is the limiting distribution of geometric distribution. Moreover, if $V\sim \msf{Exp}(\lambda)$, then $\lfloor V \rfloor$ is geometrically distributed with successful probability $1-\varepsilon=1-e^{-\lambda}$. Since $1-e^{-1/3}\approx 0.2835$, we expect the results from the two settings to be close to each other.



\begin{figure*}
\centering
\begin{subfigure}[t]{0.3\linewidth}
\centering
\includegraphics[width=2in]{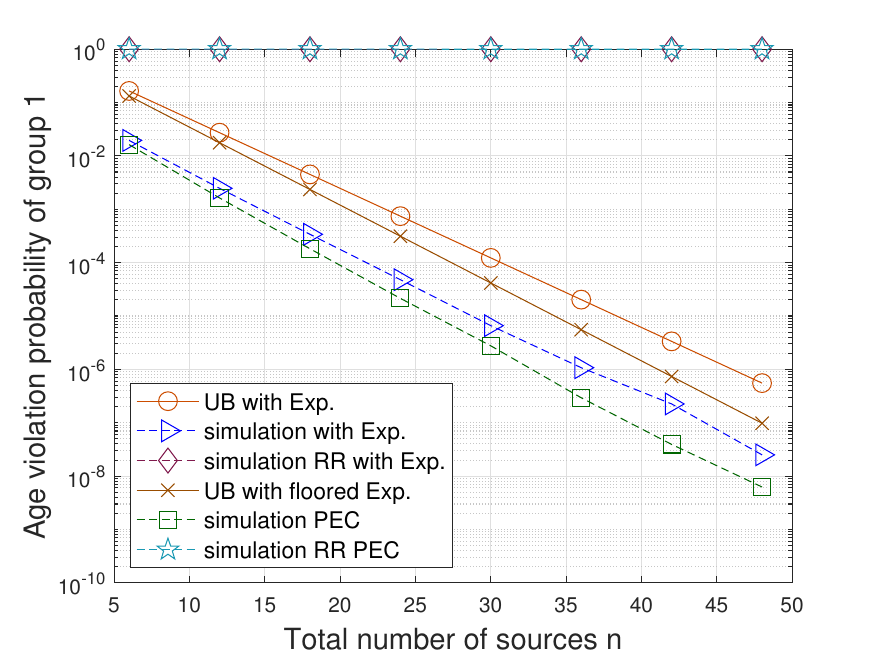}
\end{subfigure}%
\begin{subfigure}[t]{0.3\linewidth}
\centering
\includegraphics[width=2in]{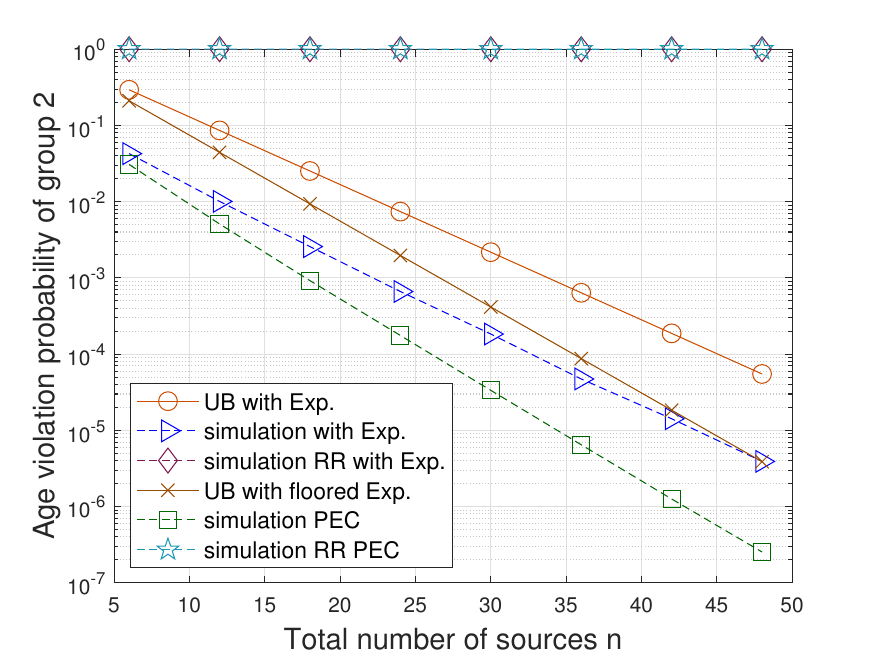}
\end{subfigure}
\begin{subfigure}[t]{0.3\linewidth}
\centering
\includegraphics[width=2in]{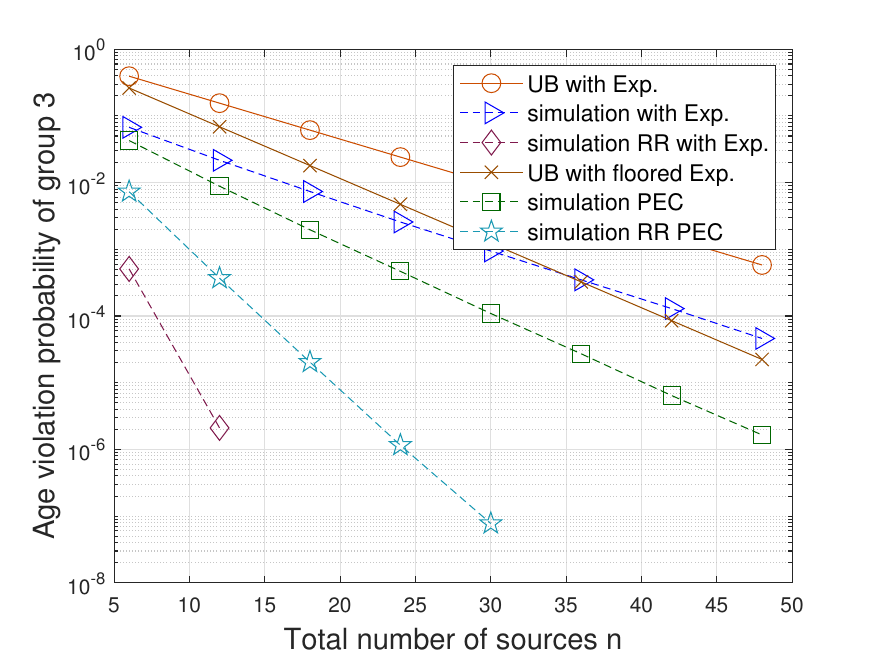}
\end{subfigure}
\caption{Age violation probability against the $n$ under the IPQ with ($x_1$,$x_2$,$x_3$)=(8,14,25).}
\label{fig:d124 increaase_n FCFS}
\end{figure*}

\begin{figure*}
\centering
\begin{subfigure}[t]{0.3\linewidth}
\includegraphics[width=2in]{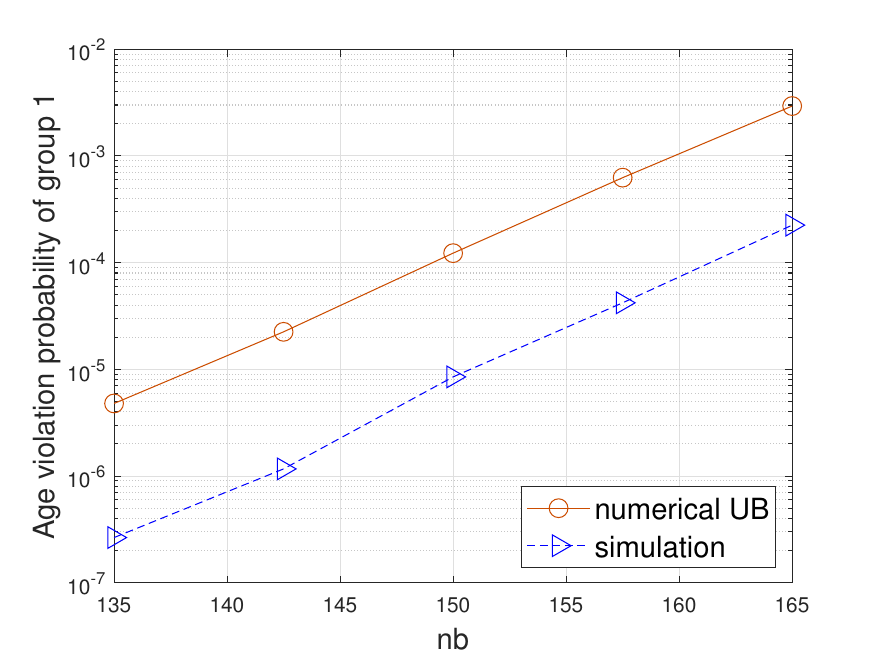}
\end{subfigure}%
\begin{subfigure}[t]{0.3\linewidth}
\includegraphics[width=2in]{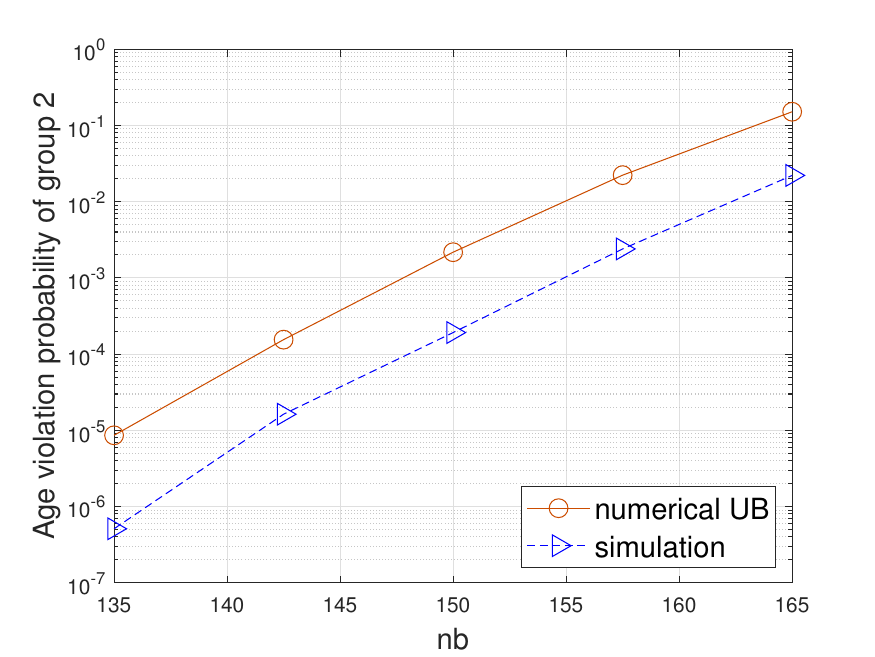}
\end{subfigure}
\begin{subfigure}[t]{0.3\linewidth}
\includegraphics[width=2in]{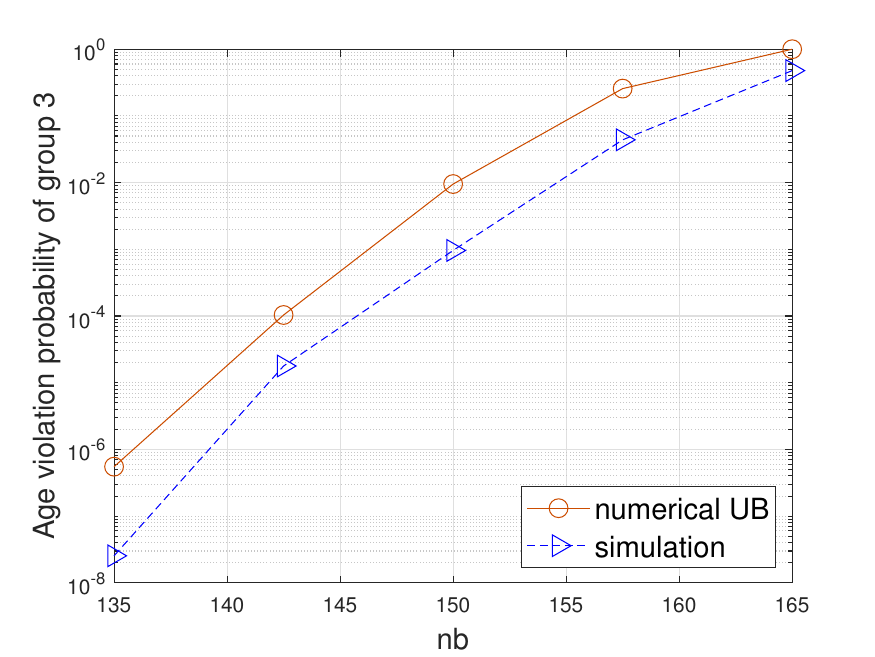}
\end{subfigure}
\caption{Age violation probability against the arrival period under the IPQ with $(x_1,x_2,x_3)=(8,14,25)$ and $n=30$.}
\label{fig:d124 increase_arr_IPQ}
\end{figure*}

\begin{figure*}
\centering
\begin{subfigure}[h]{0.3\linewidth}
\centering
\includegraphics[width=2in]{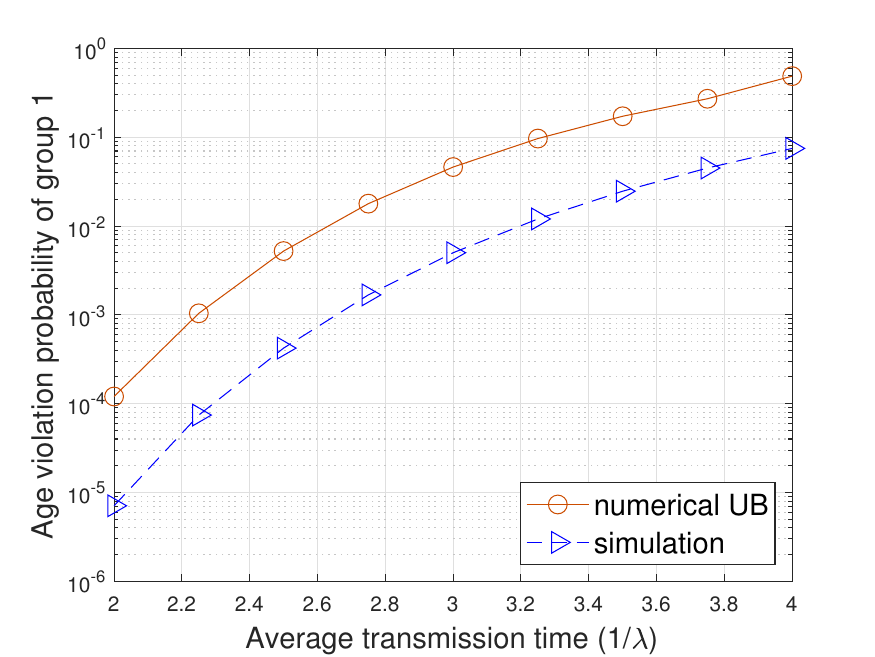}
\end{subfigure}%
\begin{subfigure}[h]{0.3\linewidth}
\centering
\includegraphics[width=2in]{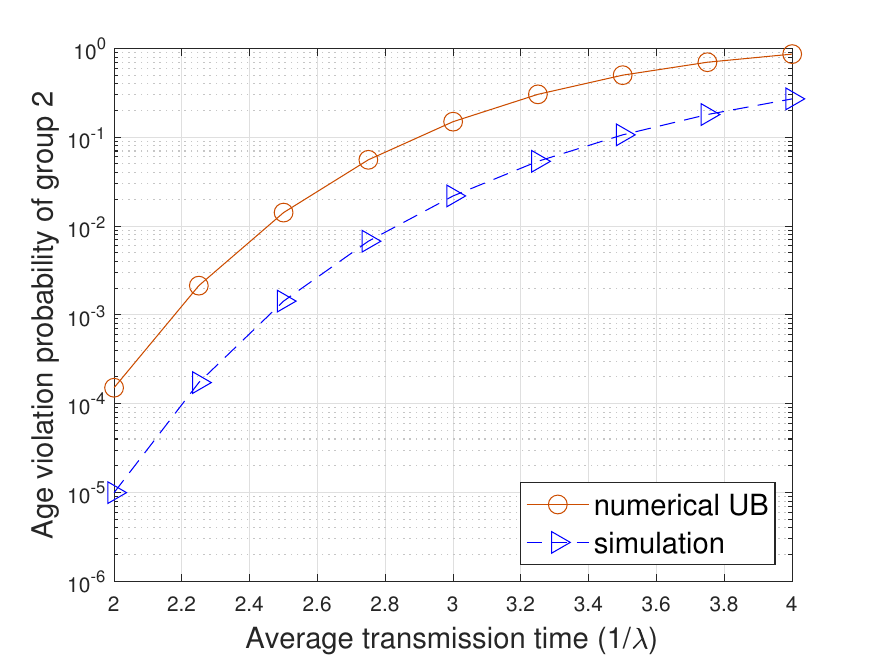}
\end{subfigure}
\begin{subfigure}[h]{0.3\linewidth}
\centering
\includegraphics[width=2in]{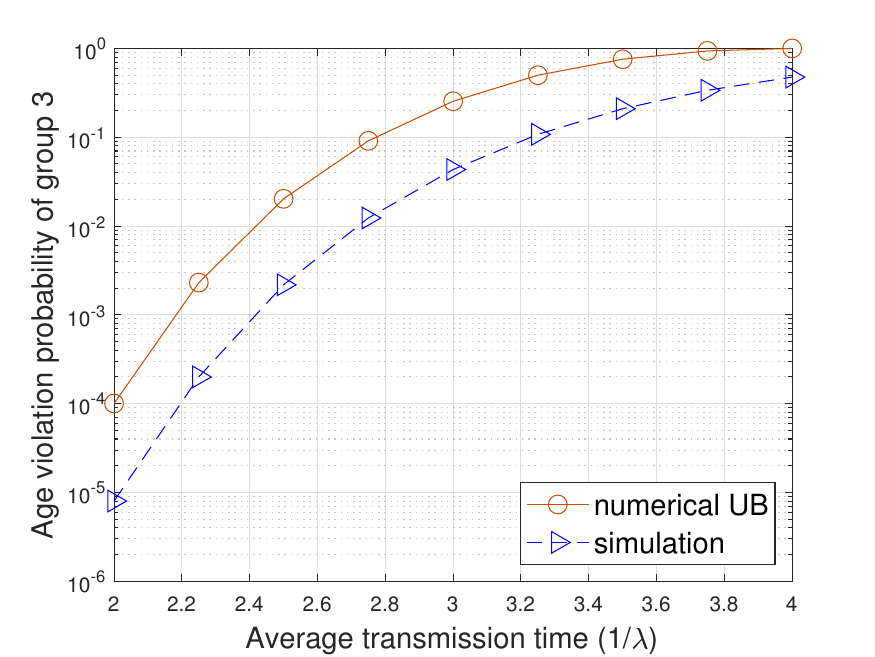}
\end{subfigure}
\caption{Age violation probability against the average transmission time under the IPQ with $ (x_1,x_2,x_3)=(7,13,24)$ and $n=30$.}
\label{fig:d137 increase_t FCFS}
\end{figure*}

In Figs. \ref{fig:d124 increaase_n FCFS}-\ref{fig:d137 increase_t FCFS}, we use computer simulation to verify the analytic results in IPQ. We set $1/\lambda=3$ and the threshold of group $g$ to be $n x_g$ with $(x_1,x_2,x_3)=(8,14,25)$. In Fig.~\ref{fig:d124 increaase_n FCFS}, the age violation probability of each group's last transmitted source against $n$ is plotted, where the upper bound \eqref{eqn:age violation prob UB} with only exponential term is also plotted. In this case, the lower bound in \eqref{eqn:age violation prob LB} with $\epsilon=0$ matches the upper bound.  One observes that the simulation and the numerical results exhibit the same slope when $n$ is large for both continuous-time and discrete-time settings, demonstrating the accuracy of our analysis. The results also confirm that as the total number of sources $n$ grows, scaling the arrival period and threshold linearly with $n$ is sufficient to drive the age violation probability vanishing. Furthermore, the comparison with the conventional RR policy is also provided, indicating that by incorporating the underlying heterogeneity into the design, the proposed GRR policy is significantly superior to RR, except for the last group of users whom the RR favors the most.
In Fig. \ref{fig:d124 increase_arr_IPQ}, we plot the age violation probability versus the arrival period by setting $(x_1,x_2,x_3)=(8,14,25)$ and $n=30$ with different $nb \in [135,165]$. The results show that the peak AoI violation probability increases as the arrival periods increase. In Fig. \ref{fig:d137 increase_t FCFS}, we consider $(x_1,x_2,x_3)=(7,13,24)$ and $n=30$ with different $1/\lambda\in [2,4]$ and validate the theoretical results under different transmission rates. We observe that the age violation probability increases as the average transmission time increases. Moreover, both Figs. \ref{fig:d124 increase_arr_IPQ} and \ref{fig:d137 increase_t FCFS} again demonstrate that our analysis well predicts the simulation results. 

\begin{figure*}
\centering
\begin{subfigure}[t]{0.3\linewidth}
\centering
\includegraphics[width=2in]{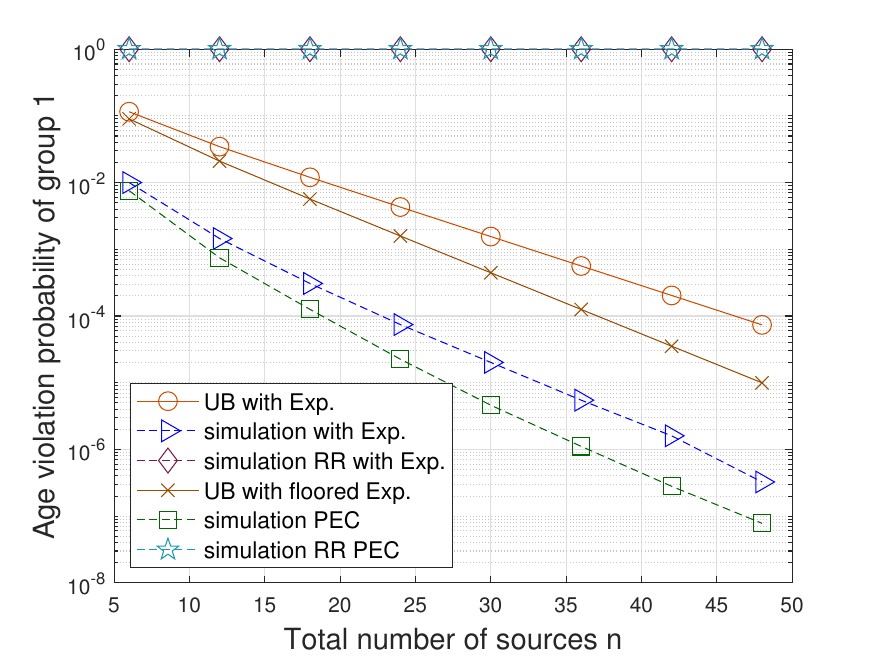}
\end{subfigure}%
\begin{subfigure}[t]{0.3\linewidth}
\centering
\includegraphics[width=2in]{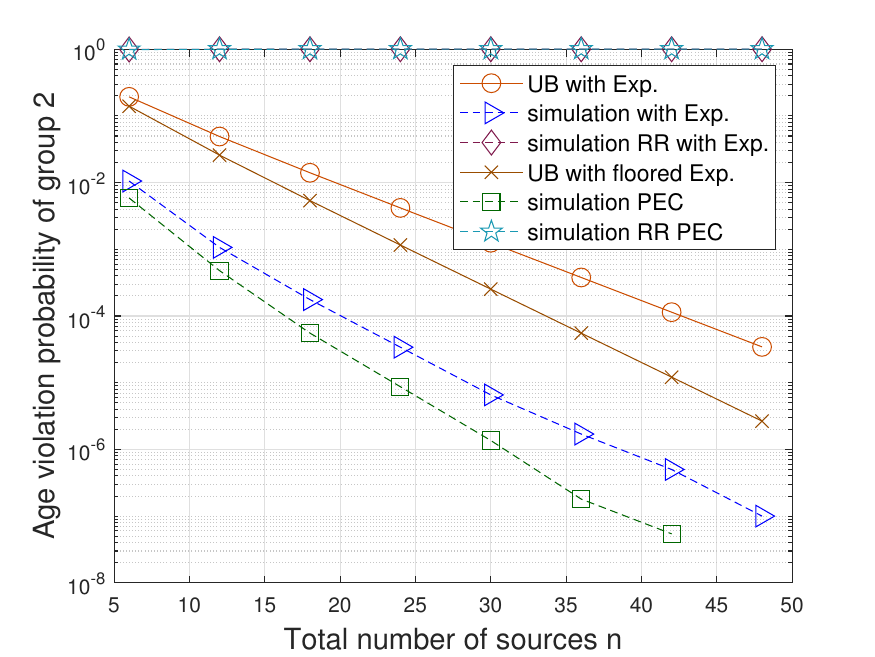}
\end{subfigure}
\begin{subfigure}[t]{0.3\linewidth}
\centering
\includegraphics[width=2in]{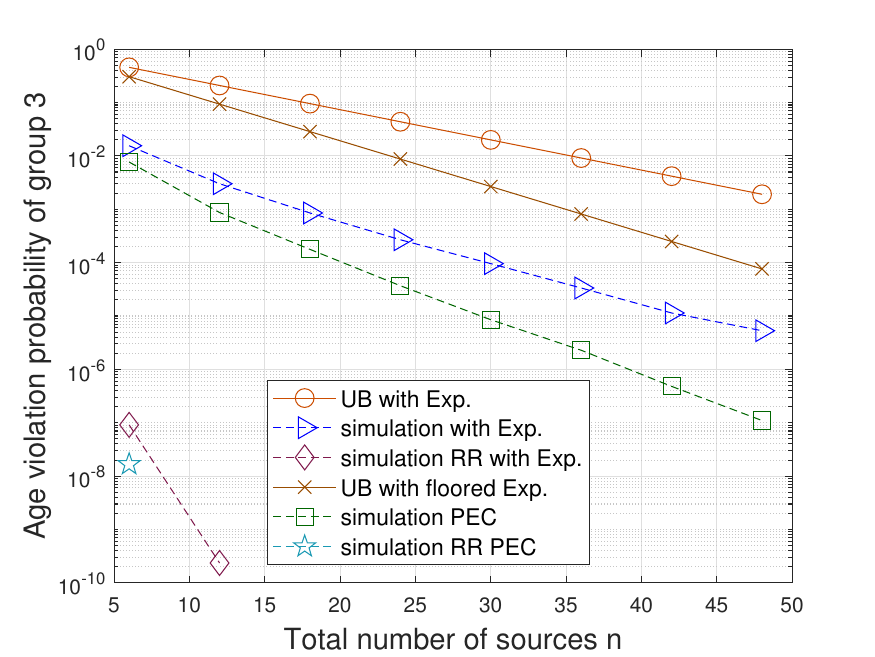}
\end{subfigure}
\caption{Age violation probability against the $n$ under the SPQ with $(x_1,x_2,x_3)=(13.5,21,36)$.}
\label{fig:d123 increase_n LCFS}
\vspace{-5pt}
\end{figure*}

\begin{figure*}
\centering
\begin{subfigure}[t]{0.3\linewidth}
\includegraphics[width=2in]{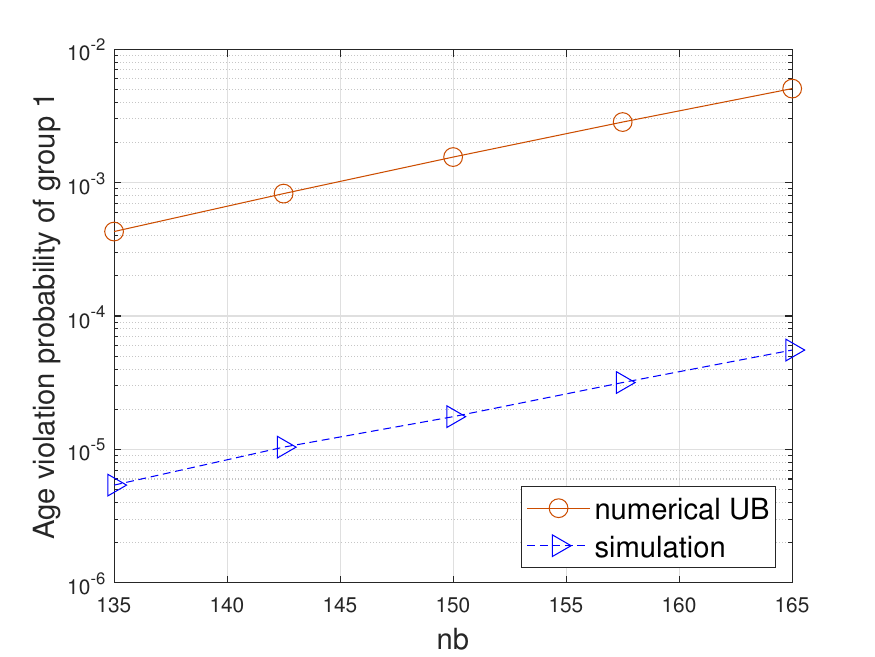}
\end{subfigure}%
\begin{subfigure}[t]{0.3\linewidth}
\includegraphics[width=2in]{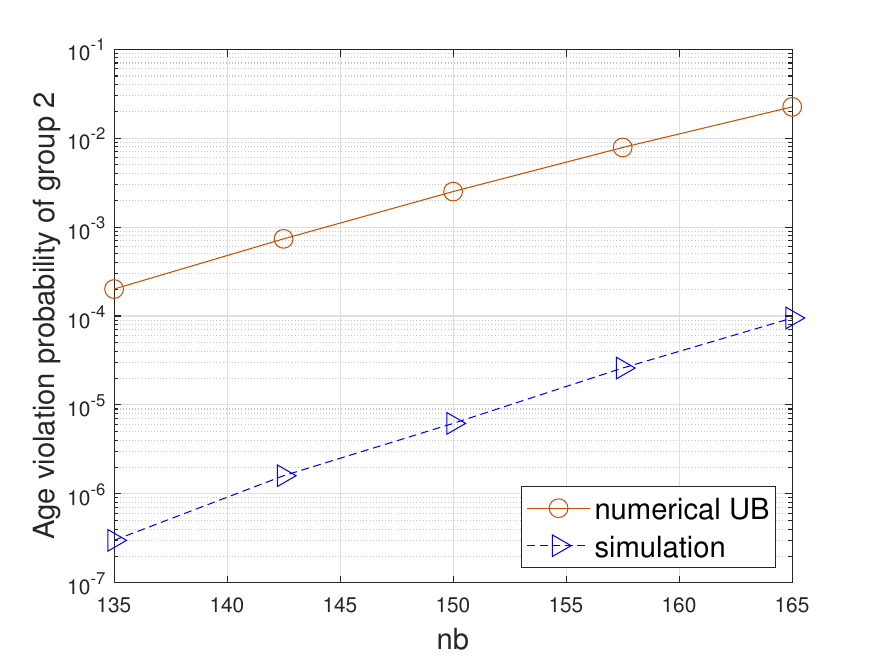}
\end{subfigure}
\begin{subfigure}[t]{0.3\linewidth}
\includegraphics[width=2in]{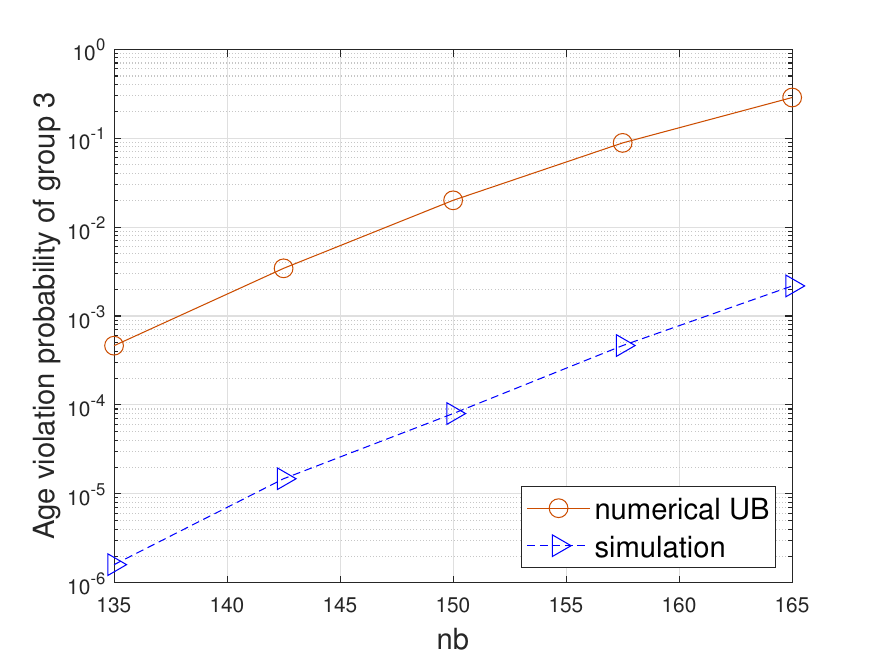}
\end{subfigure}
\caption{Age violation probability against the arrival period under the SPQ with $(x_1,x_2,x_3)=(13.5,21,36)$ and $n=30$.}
\label{fig:d124 increase_arr_LCFS}
\vspace{-5pt}
\end{figure*}

\begin{figure*}
\centering
\begin{subfigure}[t]{0.3\linewidth}
\includegraphics[width=2in]{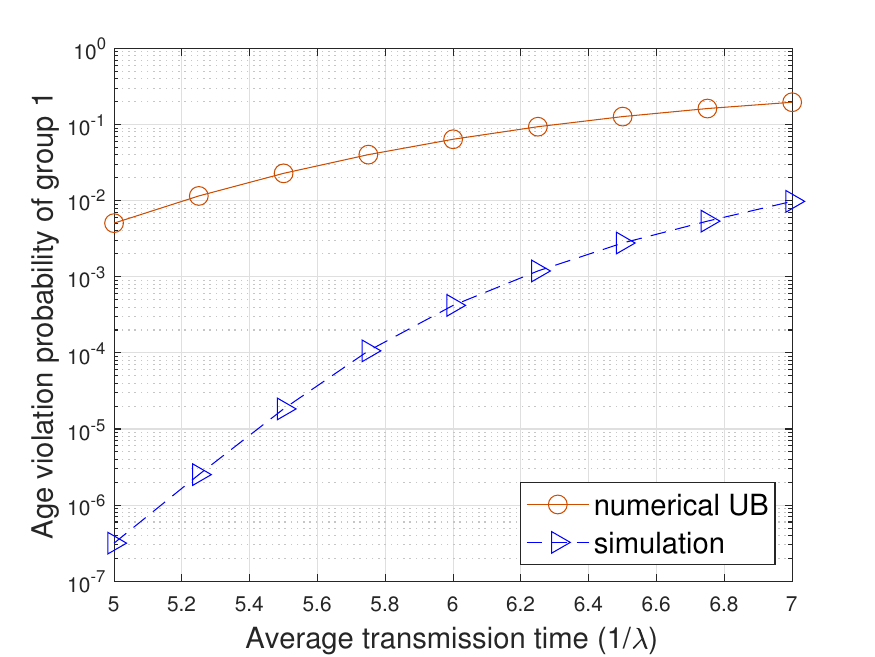}
\end{subfigure}%
\begin{subfigure}[t]{0.3\linewidth}
\includegraphics[width=2in]{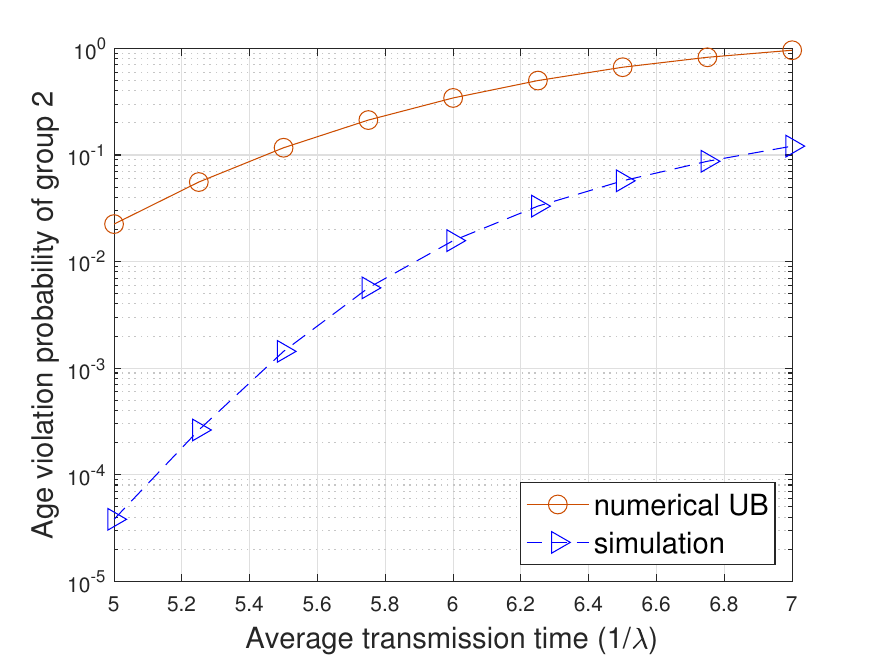}
\end{subfigure}
\begin{subfigure}[t]{0.3\linewidth}
\includegraphics[width=2in]{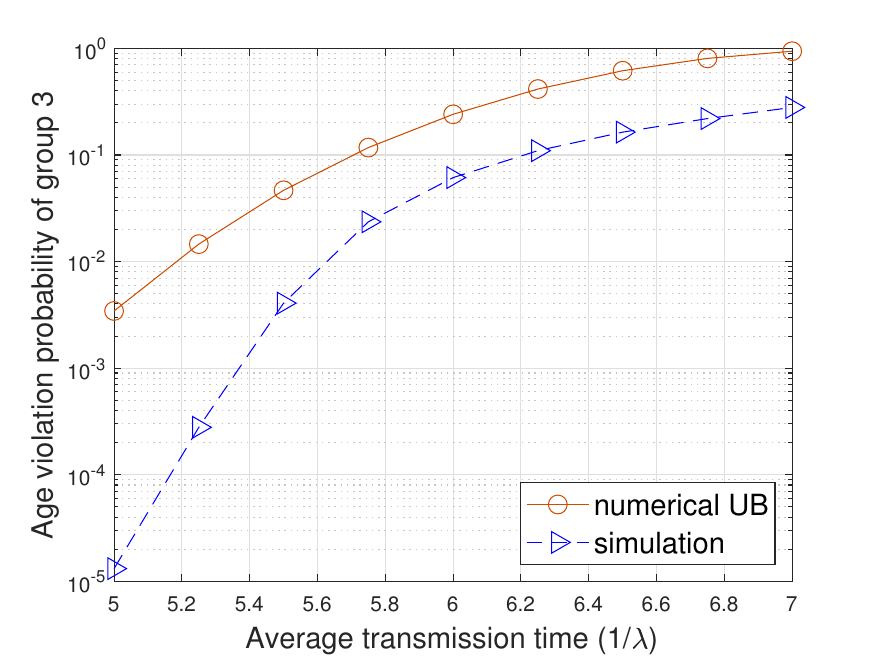}
\end{subfigure}
\caption{Age violation probability against the average transmission time under the SPQ with $(x_1,x_2,x_3)=(13,20,37)$ and $n=30$.}
\label{fig:d123 increase_t_LCFS}
\end{figure*}

We next consider SPQ in Figs.~\ref{fig:d123 increase_n LCFS}-\ref{fig:d123 increase_t_LCFS}, we set $(x_1,x_2,x_3)=(13.5,21,36)$, and $1/\lambda=5$ and plot the simulation results and analytic results in \eqref{eqn:upper_bound_preemptive}. Firstly, the results show that the simulation and the analytic results share the same slope as $n$ becomes large, which again verify the effectiveness of our analysis in both exponentially distributed and geometrically distributed (PEC channel) cases. Secondly, same conclusion regarding the linear scaling of arrival period as that in IPQ can be drawn here. Thirdly, in this case, the age violation probability under IPQ would always be 1 due to severe queue overflow caused by $1/\lambda=b=5$. This result shows that the SPQ can handle situations even when the IPQ is overflowed. Last, similar to the IPQ case, the proposed GRR significantly outperforms the conventional RR, except for the last group of sources. In Fig. \ref{fig:d124 increase_arr_LCFS}, the age violation probability of peak age versus $nb$ for $n=30$, $(d_1,d_2,d_3)=(1,2,4)$, and $(x_1,x_2,x_3)=(13.5,21,36)$ is plotted. The results demonstrate that the age violation probability increases as the arrival period increases, whose slope is captured by our analysis. In Fig. \ref{fig:d123 increase_t_LCFS}, we plot the age violation probability of peak age against $1/\lambda$ for the fixed $n=30$. We set $(d_1,d_2,d_3)=(1,2,4)$ and $(x_1,x_2,x_3)=(13,20,37)$ and consider $1/\lambda\in [5,7]$. Similar to IPQ, the age violation probability increaes as $1/\lambda$ increaes.
It is worth noting that the large gap when $1/\lambda$ is small is due to the bounding technique we used in Appendix~\ref{apx:proof_upper_bound_preemptive}. Finding a tighter bound to bridge the gap for small $1/\lambda$ is left for future investigation.



\begin{figure}
     \centering
     \begin{subfigure}[t]{0.3\linewidth}
         \centering
        \includegraphics[width=2in]{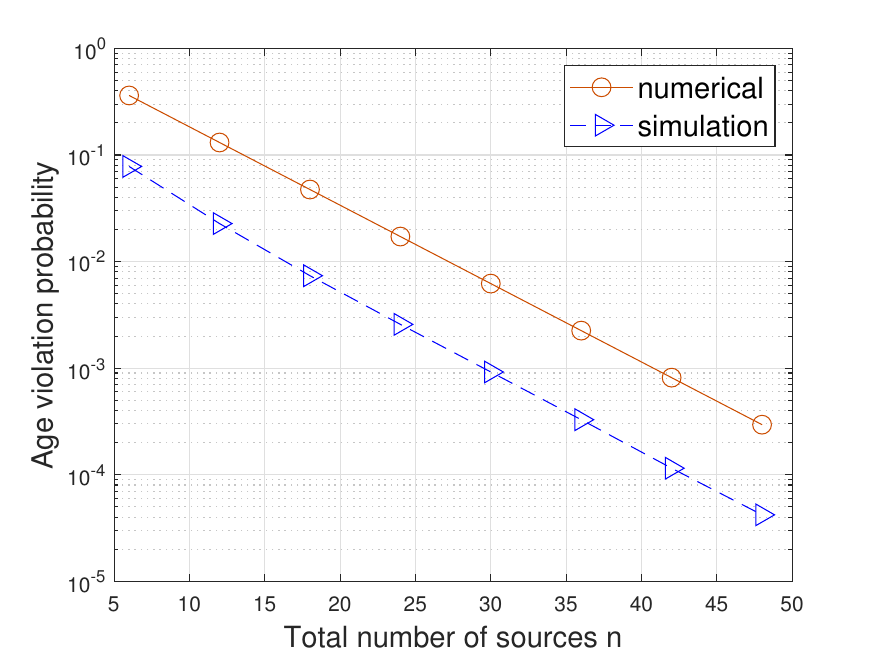}
         \caption{IPQ.}
    \label{fig:simulation_FCFS}
     \end{subfigure}
     \begin{subfigure}[t]{0.3\linewidth}
         \centering
         \includegraphics[width=2in]{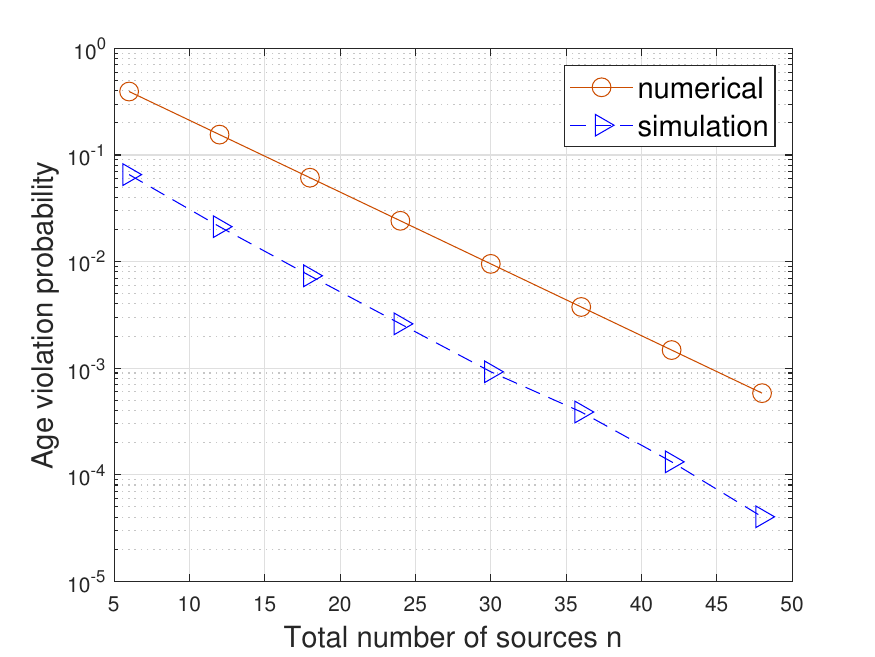}
         \caption{SPQ.}
    \label{fig:simulation_LCFS}
     \end{subfigure}
     \hfill
     \caption{Age violation probability against the $n$ for $b=5$, $x=10$, and $V_i(k)\sim \mathsf{Exp}(1/3)$.}
\end{figure}

In Fig. \ref{fig:simulation_FCFS}, we consider the homogeneous case with $b=5$, $1/\lambda = 3$ and $x=10$. The simulation results and the analytical results \eqref{eqn:upper bound_bound RR FCFS} display the same slope when $n$ is large. The gap between the two curves is due to the fact that the precise constant term is ignored in the analysis. For SPQ case, in Fig.~\ref{fig:simulation_LCFS} under the same setting as Fig.~\ref{fig:simulation_FCFS}, it is again shown that the two curves have the same slope when $n$ is large. These results confirm the effectiveness and accuracy of our analysis.

Last but not the least, we compare the performance between two different queuing disciplines in the homogeneous source system. We first focus on the large transmission rate case. In this case, the asymptotic decay rate in IPQ can be expressed as \eqref{eqn:asym_FCFS RR} with $\ell'=0$, which is given by
\begin{equation}\label{eqn:asym_prob FCFS under small transmission time}
    -\lim_{n\to \infty} \frac{1}{n} \text{Pr}(A_i(k) \geq nx) = \sup_{\theta}\left\{ \theta x - \theta b - \Lambda_v(\theta) \right\}.
\end{equation}
For SPQ, by plugging \eqref{eqn:rate_function_single_packet_infty} into \eqref{eqn:asym_prob_single_packet_RR}, we arrive at the same expression as \eqref{eqn:asym_prob FCFS under small transmission time}. This hints that the two queueing disciplines would achieve the same asymptotic age violation performance under this condition. This is precisely what can be observed in Figs. \ref{fig:simulation_FCFS} and \ref{fig:simulation_LCFS}, where the age violation probabilities for the two different queuing systems are almost the same. On the other hand, for a general case, since the minimizer in \eqref{eqn:asym_FCFS RR} may not be $\ell'=0$, our analysis concretizes the intuition that SPQ would result in better performance than IPQ.

\section{Conclusion}\label{sec:conclusion}
In this work, we considered status update in the heterogeneous multi-source system where multiple groups of sources wish to maintain the freshness of their information. The GRR scheduling policy was proposed and its AoI performance guarantee under the IPQ and that under the SPQ were investigated, which have enabled us to derive the asymptotic decay rate in the multi-source regime to capture the AoI behavior in the many-source regime. The results indicated that the proposed GRR policy is capable of handling many source types with heterogeneous arrivals and leads to larger age violation probability decay rates for types of sources whose packets arrive more frequently. Moreover, we showed that with the proposed GRR policy, when the total number of sources $n$ grows, scaling the arrival period linearly with $n$ suffices to drive the age violation probability vanishing. Extensive simulations were conducted to validate our analytic results. The results showed that the proposed GRR policy significantly outperforms the conventional RR policy. The simulation results also confirmed that with GRR the SPQ outperforms the IPQ in terms of the age violation probability even in situations that IPQ is overflowed. When specialized to the homogeneous case, our analysis unveiled a surprising fact that the two disciplines lead to the same asymptotic decay rate when the inter-arrival time is much larger than the total transmission time. A potential future research direction is to extend the proposed policy and analytical framework to systems that enable parallel transmissions.

\begin{appendices}

\section{Proof of Lemma~\ref{lemma:A(k) iterative formulation}}\label{apx:proof_peak_age_formulation_FCFS}
\begin{IEEEproof}
     Substituting $W_{1,1}(\Tilde{k})$ in \eqref{eqn:A(k) FCFS} with \eqref{eqn:W(k) FCFS} yields
	\begin{align}
	&A_{g,i}(k) = \max \left\{ \vphantom{\sum_{(g,u,j)\in \mc{J}_{g,i}(k)} V_{g,u}(j)} W_{1,1}(\Tilde{k}-1)+V(\Tilde{k}-1) - nb,\ 0\right\} + \sum_{(g',i',j)\in \mc{J}^-_{g,i}(k)} V_{g',i'}(j) + d_g nb \nonumber \\
	&\overset{(a)}{=} \max \left\{  \left( \vphantom{\sum_{(g,u,j)\in \mc{J}_{g,i}(k)} V_{g,u}(j)} W_{1,1}(\Tilde{k}-2) + V(\Tilde{k}-2) - nb\right)^+ +V(\Tilde{k}-1) - nb,\ 0\right\} + \sum_{(g',i',j)\in \mc{J}^-_{g,i}(k)} V_{g',i'}(j) + d_g nb \nonumber \\
	&= \max \left\{ \vphantom{\sum_{(g,u,j)\in \mc{J}_{g,i}(k)} V_{g,u}(j)} W_{1,1}(\Tilde{k}-2) + \sum_{r=\Tilde{k}-2}^{\Tilde{k}-1} V(r) - 2nb,\  V(\Tilde{k}-1) - nb,\ 0\right\} + \sum_{(g',i',j)\in \mc{J}^-_{g,i}(k)} V_{g',i'}(j) + d_g nb \nonumber \\
	&= \max \left\{ \vphantom{\sum_{(g,u,j)\in \mc{J}_{g,i}(k)} V_{g,u}(j)} W_{1,1}(\Tilde{k}-3) + \sum_{r=\Tilde{k}-3}^{\Tilde{k}-1} V(r)-3nb,\  \sum_{r=\Tilde{k}-2}^{\Tilde{k}-1} V(r) - 2nb,\  V(\Tilde{k}-1) - nb,\ 0\right\}\nonumber\\
	&\qquad + \sum_{(g',i',j)\in \mc{J}^-_{g,i}(k)} V_{g',i'}(j) + d_g nb \nonumber \\
	&=\dots= \max_{1\leq \ell \leq \Tilde{k}} \left\{ \sum_{r=\ell}^{\Tilde{k}-1} V(r) - (\Tilde{k}-\ell)nb \right\} + \sum_{(g',i',j)\in \mc{J}^-_{g,i}(k)} V_{g',i'}(j) + d_g nb. \label{eqn:proof peak_age_expression}
	\end{align}
This complete the proof.
\end{IEEEproof}

\section{Proof of Theorem~\ref{thm:upper_bound_FCFS}}\label{apx:proof_upper_bound_FCFS}
\begin{IEEEproof}
    From the age evolution derived in  \eqref{eqn:A(k) iterative formulation}, we have
	\begin{align}
	\text{Pr} \left( \vphantom{\sum_{r=s}^{k-1} \sum_{u=1}^n V_u(r)} A_{g,i}(k)\geq nx\right) &= \text{Pr} \left( \max_{1\leq \ell\leq \Tilde{k}} \left[ \sum_{r=\ell}^{\Tilde{k}-1} V(r) - (\Tilde{k}-\ell) nb \right] + \sum_{(g',i',j)\in \mc{J}^-_{g,i}(k)} V_{g',i'}(j) + d_g nb \geq nx \right)\nonumber\\
	&\overset{(a)}{\leq} \sum_{\ell=1}^{\Tilde{k}} \text{Pr} \left( \sum_{r=\ell}^{\Tilde{k}-1} V(r) - (\Tilde{k}-\ell) nb + \sum_{(g',i',j)\in \mc{J}^-_{g,i}(k)} V_{g',i'}(j) + d_g nb \geq nx \right)\nonumber\\
	&\overset{(b)}{\leq} \sum_{\ell=1}^{\Tilde{k}}  \mathbb{E}\left[ e^{\theta \left( \sum_{r=\ell}^{\Tilde{k}-1} V(r) + \sum_{(g',i',j) \in \mc{J}^-_{g,i}(k)} V_{g',i'}(j) \right) } \right] e^{-(\Tilde{k}-\ell-d_g) n\theta b} e^{-n\theta x},\label{eqn:derive Pr(A>nx) part 1}
	\end{align}
	where (a) follows from the union bound, and (b) uses the Chernoff bound along for a constant $\theta > 0$. 
    Note that the expectation term in \eqref{eqn:derive Pr(A>nx) part 1} can be further bounded as
    \begin{align}
	&\mathbb{E}\left[e^{\theta \left(\sum_{r=\ell}^{\Tilde{k}-1}V(r) + \sum_{(g',i',j)\in \mc{J}^-_{g,i}(k) }V_{g',i'}(j)\right)}\right]
    \overset{(c)}{=} \prod_{r=\ell}^{\Tilde{k}-1} \mathbb{E}\left[e^{\theta V(r)} \right] \prod_{(g',i',j)\in \mc{J}^-_{g,i}(k)} \mathbb{E} \left[ e^{\theta V_{g',i'}(j)}\right]\nonumber\\ 
    &\overset{(d)}{=} e^{\sum_{r=\ell}^{\Tilde{k}-1}|\mc{J}^-_{g^*,n_{g^*}}(r)|\Lambda_v(\theta)} e^{n m_{g,i}(k) \Lambda_v(\theta)}
    \overset{(e)}{\leq}  e^{\ell' \sum_{j=1}^\eta \frac{\Tilde{d}}{d_j}n_j \Lambda_v(\theta)} e^{n m_{g,i}(k) \Lambda_v(\theta)}\label{eqn:iterative V UB},
    \end{align}
    where (c) is due to the fact that $V_{g,i}(k)$s are independent, (d) follows from that $V_{g,i}(k)$s are identically distributed and source $(g^*, n_{g^*})$ is the last transmitted source in round $r$. (e) is because $\sum_{r=\ell}^{\tilde{k}-1}|\mc{J}^-_{g^*,n_{g^*}}(r)|$ is the total number of transmission packets of any sources from round $r$ to round $\tilde{k}-1$ is no larger than $\ell'\sum_{j=1}^{\eta} \frac{\Tilde{d}}{d_j}n_j$ the number of updates in $\ell'=\lceil \frac{\Tilde{k}-\ell}{\Tilde{d}} \rceil$ iterations. Plugging \eqref{eqn:iterative V UB} into \eqref{eqn:derive Pr(A>nx) part 1} shows that
	\begin{align}
	&\eqref{eqn:derive Pr(A>nx) part 1} \leq  \sum_{\ell=1}^{\Tilde{k}}  e^{ \ell'  \sum_{j=1}^\eta \frac{\Tilde{d}}{d_j}n_j \Lambda_v(\theta)} e^{-(\ell'-1)\Tilde{d} n \theta b} e^{n m_{g,i}(k) \Lambda_v(\theta)} e^{d_g n \theta b} e^{-n \theta x}  \nonumber\\
	&\overset{(f)}{\leq} \tilde{d} \sum_{\ell'=1}^{\lceil\frac{\Tilde{k}-1}{\Tilde{d}}\rceil}  e^{ -n \ell' \left( \Tilde{d} \theta b - \sum_{j=1}^\eta \frac{\Tilde{d}}{d_j}\alpha_j \Lambda_v(\theta) \right)} e^{n(\Tilde{d}+d_g) \theta b} e^{n m_{g,i}(k) \Lambda_v(\theta)} e^{-n \theta x} + e^{n d_g \theta b} e^{n m_{g,i}(k)\Lambda_v(\theta)}e^{-n\theta x}\nonumber\\
    &\leq \Tilde{d} \sum_{\ell'=1}^\infty e^{ -n \ell' \left( \Tilde{d} \theta b - \sum_{j=1}^\eta \frac{\Tilde{d}}{d_j}\alpha_j \Lambda_v(\theta) \right)} e^{n(\Tilde{d}+d_g) \theta b} e^{n m_{g,i}(k) \Lambda_v(\theta)} e^{-n \theta x} + e^{n d_g \theta b} e^{n m_{g,i}(k)\Lambda_v(\theta)}e^{-n\theta x}\nonumber \\ 
    &\overset{(g)}{\leq} \Tilde{d} \sum_{\ell'=1}^{c'-1} e^{-n \theta x - \left( \left( \ell' - 1 \right) \Tilde{d} - d_g \right) n\theta b - \left( \ell' \sum_{j=1}^\eta \frac{\Tilde{d}}{d_j}\alpha_j + m_{g,i}(k) \right) \Lambda_v(\theta) }\nonumber\\
    &+ \Tilde{d} \sum_{\ell'=c'}^\infty e^{ -n \ell' \left( \Tilde{d} \theta b - \sum_{j=1}^\eta \frac{\Tilde{d}}{d_j}\alpha_j \Lambda_v(\theta) \right)} e^{n(\Tilde{d}+d_g) \theta b} e^{n m_{g,i}(k) \Lambda_v(\theta)} e^{-n \theta x} + e^{n d_g \theta b} e^{n m_{g,i}(k)\Lambda_v(\theta)}e^{-n\theta x}, \label{eqn:derive Pr(A>nx) part 2}
	\end{align}
	where in (f) we sum over $\ell'$ instead of $\ell$ and get the upper bound because there are at most $\tilde{d}$ different $\ell$ matching to the same $\ell'$. In (g), we split the infinite series into two terms with $\ell'<c'$ and $\ell'\geq c'$, respectively.\footnote{Note that the infimum of $\sup_{\theta}\{-n\ell'\left( \Tilde{d} \theta b - \sum_{j=1}^\eta \frac{\Tilde{d}}{d_j}\alpha_j \Lambda_v(\theta) \right)\}$ is achieved for some finite $\ell'$, we pick a $c'$ so that it is achieved at $\ell'<c'$.} Since \eqref{eqn:derive Pr(A>nx) part 2} holds for every $\theta>0$, we take the best one for the first term in what follows:
 
 
    \begin{align}
    &\text{Pr}(A_{g,i}(k) \geq nx)\leq \Tilde{d} \sum_{\ell'=1}^{c'-1} e^{-n \ell' I_{g,i}^{(U,\ell')}\left(\frac{x}{\ell'} + \frac{(\ell'-1)-d_g}{\ell'}b,k,n\right)} \nonumber\\
    &\quad + \Tilde{d}  \frac{e^{-n  c' \left( \Tilde{d}\theta b - \sum_{j=1}^\eta \frac{\Tilde{d}}{d_j} \alpha_j \Lambda_v(\theta) \right)}}{1-e^{-n \left( \Tilde{d}\theta b - \sum_{j=1}^\eta \frac{\Tilde{d}}{d_j} \alpha_j \Lambda_v(\theta)\right)}} e^{n m_{g,i}(k) \Lambda_v(\theta)}  e^{n (\Tilde{d} + d_g) \theta b} e^{-n\theta x} + e^{n I_{g,i}^{(U,k)}\left(x-d_g b,k,n\right)}  \nonumber\\
    &\leq \Tilde{d} \sum_{\ell'=1}^{c'-1} e^{-n \ell' I_{g,i}^{(U,\ell')}\left( \frac{x}{\ell'}+\frac{(\ell'-1)\Tilde{d}-d_g}{\ell'}b,k,n \right)} \nonumber\\
    &\quad + \Tilde{d}  e^{-n  c' \left( \Tilde{d}\theta b - \sum_{j=1}^\eta \frac{\Tilde{d}}{d_j} \alpha_j \Lambda_v(\theta) \right)} e^{n m_{g,i}(k) \Lambda_v(\theta)}  e^{n (\Tilde{d} + d_g) \theta b} e^{-n\theta x} + e^{n I_{g,i}^{(U,k)}\left(x-d_g b,k,n\right)} \nonumber\\
    &\leq (\Tilde{d} c'+1) \cdot e^{-n \min_{0\leq \ell' \leq k'} \gamma_{g,i}^{(U,\ell')}(x,k,n)},
    \end{align}
which completes the proof.
\end{IEEEproof}

\section{Proof of Theorem~\ref{thm:lower_bound_FCFS}}\label{apx:proof_lower_bound_FCFS}
We note analyzing \eqref{eqn:A(k) iterative formulation} with any particular $\ell$ would lead to a lower bound on the age violation probability of peak age. One can then choose the $\ell$ that results in the largest lower bound. In this appendix, we split the proof into two cases, namely $\ell\neq \tilde{k}$ and $\ell=\tilde{k}$. 

\begin{IEEEproof}

\underline{Case 1 ($1\leq \ell \leq \Tilde{k}-1$ or equivalently $1 \leq \ell' \leq k'$):}
\begin{align}
    \text{Pr} \left( \vphantom{\sum_{r=s}^{k-1} \sum_{u=1}^n V_u(r)} A_{g,i}(k)\geq nx\right) &= \text{Pr} \left( \max_{1\leq \ell\leq \Tilde{k}} \left[ \sum_{r=\ell}^{\Tilde{k}-1} V(r) - (\Tilde{k}-\ell) nb \right] + \sum_{(g',i',j)\in \mc{J}^-_{g,i}(k)} V_{g',i'}(j) + d_g nb \geq nx \right)\nonumber\\
    &\overset{(a)}{\geq} Pr\left( \sum_{r=\ell}^{\Tilde{k}-1} V(r) - (\Tilde{k}-\ell) nb + \sum_{(g',i',j)\in \mc{J}^-_{g,i}(k)} V_{g',i'}(j) + d_g nb \geq nx \right)\nonumber\\
    &\overset{(b)}{\geq}  \mathbb{E}\left[ e^{\theta\left(\sum_{r=\ell}^{\Tilde{k}-1}V(r) + \sum_{(g',i',j) \in \mc{J}_{g,i}(k)} V_{g',i'}(j)\right)} \right]  e^{-n (\Tilde{k}-\ell) \theta b} e^{n d_g \theta b} e^{-n \theta x} e^{-n \epsilon}\nonumber \\
    &\overset{(c)}{\geq} e^{\left( \vphantom{\sum_{r=1}^{(\Tilde{d}/d_g)}}\ell'-1 \right) \sum_{j=1}^\eta \frac{\Tilde{d}}{d_j}n_j \Lambda_v(\theta)}  e^{n m_{g,i}(k) \Lambda_v(\theta)} e^{-n \left(\ell' \Tilde{d} - d_g \right) \theta b}   e^{-n \theta x} e^{-n \epsilon}\nonumber\\
	&\geq e^{-n \left\{ \ell' I_{g,i}^{(L,\ell')}\left( \frac{x}{\ell'}+\frac{\ell'\Tilde{d}-d_g}{\ell'}b,k,n\right) + \epsilon \right\} } = e^{-n \left( \gamma_{g,i}^{(L,\ell')}(x,k,n) + \epsilon \right)},
    \label{eqn:Pr(A(k)>nx) part 1}
	\end{align}
    where (a) is obtained by choosing an arbitray $1\leq \ell \leq \Tilde{k}-1$, (b) follows from the Cramer-Chernoff theorem which holds for any $\theta > 0$ and $\epsilon > 0$, and (c) can be obtained by a lower bound on the expectation term derived in a similar fashion as \eqref{eqn:iterative V UB} but with only the $\ell'-1$ iterations.

\underline{Case 2 ($\ell=\Tilde{k}$ or equivalently $\ell'=0$):}
\begin{align}
    &Pr\left( A_{g,i}(k) \geq nx \right)\overset{(a)}{=} Pr\left( \sum_{(g',i',j)\in \mc{J}^-_{g,i}(k)} V_{g',i'}(j) + d_g nb \geq nx \right)\nonumber\\
    &\overset{(b)}{\geq}  \mathbb{E}\left[ e^{\theta \sum_{(g',i',j) \in \mc{J}^-_{g,i}(k)} V_{g',i'}(j)} \right] e^{n d_g \theta b} e^{-n \theta x} e^{-n \epsilon}
    = e^{-n \left( \gamma_{g,i}^{(L,\ell'=0)}(x,k,n) + \epsilon \right)},
\end{align}
where (a) holds because $\ell=\tilde{k}$ and (b) again applies the Cramer-Chernoff bound for some $\epsilon>0$.

Combining the two cases completes the proof.
\end{IEEEproof}

\section{Proof of Corollary \ref{coro:long-run fraction of violation probability}}\label{apx:proof_of_long-run probability}
\begin{IEEEproof}
For any $k$, define $\zeta = (k \mmod \tilde{d}/d_g)+1$. The long-run fraction of violation probability is given by
    \begin{align*}
        &\lim_{\kappa\to \infty} \frac{1}{\kappa} \sum_{k=1}^\kappa \mathbb{E} \left[ \mathbbm{1}_{\{A_{g,i}(k)\geq nx\}}\right] =\lim_{\kappa\to \infty} \frac{1}{\kappa} \sum_{k=1}^\kappa \text{Pr}(A_{g,i}(k)\geq nx)\\
        &= \lim_{\kappa\to \infty} \frac{1}{\kappa}  \left\{ \sum_{k=1,\zeta=1}^\kappa \text{Pr}(A_{g,i}(k)\geq nx) + ... + \sum_{k=\Tilde{d}/d_g,\zeta=\tilde{d}/d_g}^\kappa \text{Pr}(A_{g,i}(k)\geq nx) \right\} \nonumber \\
        &\overset{(a)}{\leq} \lim_{\kappa\to \infty}\frac{1}{\kappa} \kappa \frac{d_g}{\Tilde{d}} (\Tilde{d} c'+1) e^{-n \min_{\ell' \geq 0} \gamma_{g,i}^{(U,\ell')}(x,\zeta=1,\infty)} + \lim_{\kappa\to \infty}\frac{1}{\kappa} \kappa \frac{d_g}{\Tilde{d}} (\Tilde{d} c'+1) e^{-n \min_{\ell' \geq 0} \gamma_{g,i}^{(U,\ell')}(x,\zeta=2,\infty)} \nonumber\\
        &\qquad + ... + \lim_{\kappa\to \infty}\frac{1}{\kappa} \kappa \frac{d_g}{\Tilde{d}} (\Tilde{d} c'+1) e^{-n \min_{\ell' \geq 0} \gamma_{g,i}^{(U,\ell')}(x,\zeta=\Tilde{d}/d_g,\infty)}  \\
        &=\sum_{\zeta=1}^{\Tilde{d}/d_g} \frac{d_g}{\Tilde{d}} (\Tilde{d}c'+1) e^{-n \min_{\ell' \geq 0}\gamma_{g,i}^{(U,\ell')}(x,\zeta,\infty)},
    \end{align*}
where (a) follows from \eqref{eqn:age violation prob UB} and includes all nature number into the minimization.
\end{IEEEproof}

\section{Proof of Lemma~\ref{lemma:peak_age_formulation_preemption}}\label{apx:proof_peak_age_formulation_LCFS}
We start from \eqref{eqn:A(k)} and lay out the recursion as follows:
\begin{align}
    A_{g,i}(k) &\overset{(a)}{=} \max\left\{ W_{g,i}(k-1) + \sum_{(g',i',j) \in \mc{J}^+_{g,i}(k-1)} V_{g',i'}(j), nb\right\} + V(d_g(k-2)+2) + N_{1,1}(d_g(k-2)+2) \nonumber\\
    &\quad + \dots + V(d_g(k-1)) + N_{1,1}(d_g(k-1)) + \sum_{(g',i',j)\in \mc{J}^-_{g,i}(k)}V_{g',i'}(j)\nonumber\\
    &= \max\left\{ W_{g,i}(k-1) + \sum_{(g',i',j) \in \mc{J}^+_{g,i}(k-1)} V_{g',i'}(j) + V(d_g(k-2)+2), nb+ V(d_g(k-2)+2), 2nb\right\}\nonumber\\
    &\quad + \dots + V(d_g(k-1)) + N_{1,1}(d_g(k-1)) + \sum_{(g',i',j)\in \mc{J}^-_{g,i}(k)}V_{g',i'}(j)\nonumber\\
    &=\dots= \max \left\{ W_{g,i}(k-1) + \sum_{(g',i',j) \in \mc{J}^+_{g,i}(k-1)} V_{g',i'}(j) + \sum_{r=d_g(k-2)+2}^{d_g(k-1)} V(r), \right. \nonumber\\
    &\qquad \qquad \left. \max_{2 \leq \ell \leq d_g+1} \left[ \sum_{r=d_g(k-2)+\ell}^{d_g(k-1)} V(r) + (\ell-1) nb \right] \right\} + \sum_{(g',i',j) \in \mc{J}^-_{g,i}(k)} V_{g',i'}(j),
\end{align}
where the maximization in (a) is due to the possible idle time before transmitting the $(d_g(k-2)+2)$-th update of source $(1,1)$.

\section{Proof of Theorem~\ref{thm:upper_bound_preemptive}}\label{apx:proof_upper_bound_preemptive}
\begin{IEEEproof}
Following from Lemma~\ref{lemma:peak_age_formulation_preemption}, we have
    \begin{align}
    &A_{g,i}(k) \overset{(a)}{<} \max \left\{ d_g nb + \sum_{(g',i',j) \in \mc{J}^+_{g,i}(k-1)} V_{g',i'}(j) + \sum_{r=d_g(k-2)+2}^{d_g(k-1)} V(r), \right. \nonumber\\
    &\qquad \qquad \qquad \left. \max_{2 \leq \ell \leq d_g+1} \left[ \sum_{r=d_g(k-2)+\ell}^{d_g(k-1)} V(r) + (\ell-1) nb \right] \right\} + \sum_{(g',i',j) \in \mc{J}^-_{g,i}(k)} V_{g',i'}(j)\nonumber\\
    &\overset{(b)}{=} \max \left\{ \vphantom{\sum_{r=d_g(k-2)+2}^{d_g(k-1)} V(r)} d_g nb + T_{g,i}(k-1) + V_{g,i}(k),  \max_{2 \leq \ell \leq d_g+1} \left[ \sum_{r=d_g(k-2)+\ell}^{d_g(k-1)} V(r) + (\ell-1) nb \right] + \sum_{(g',i',j) \in \mc{J}^-_{g,i}(k)} V_{g',i'}(j)\right\} 
    \nonumber\\
    &\overset{(c)}{\leq} d_g nb + T_{g,i}(k-1) + V_{g,i}(k),\label{eqn:peak_age_upper_bound_preemptive}
    \end{align}
where (a) uses $W_{g,i}(k-1) < d_g nb$ and (b) follows from the definition of $T_{g,i}(k-1)$. For (c), note that the first term of the maximization in (b) must be larger than the second term because for any $\ell'$, we have $\sum_{r=d_g(k-2)+\ell'}^{d_g(k-1)} V(r) + \sum_{(g',i',j) \in \mc{J}^-_{g,i}(k)} V_{g',i'}(j) < T_{g,i}(k-1)$ and $(\ell'-1)nb < d_g nb$. With \eqref{eqn:peak_age_upper_bound_preemptive}, we thus have
	\begin{align}
	&\text{Pr} \left( A_{g,i}(k) \geq nx \right) \leq \text{Pr} \left( d_g    nb + T_{g,i}(k-1) + V_{g,i}(k) \geq nx \right) \nonumber\\
	&\overset{(d)}{\leq} \mathbb{E}\left[ \vphantom{\sum_{r=s}^{k-1} T(r)} e^{\theta T_{g,i}(k-1) + \theta    V_{g,i}(k)} \right]    e^{n   d_g    \theta b}    e^{-n\theta x} 
	= e^{-n \left( \theta x - d_g   \theta b - \left( t_{g,i}(k)+\frac{1}{n} \right)    \Lambda_v(\theta) \right) }\label{eqn:proof single packet UB part 1},
    \end{align} 
where (d) follows from Chernoff bound. Since \eqref{eqn:proof single packet UB part 1} holds for every $\theta > 0$, picking the best $\theta$ completes the proof. 
\end{IEEEproof}

\end{appendices} 

\bibliographystyle{IEEEtran}
\bibliography{main.bbl}
\end{document}